\documentclass[onecolumn]{IEEEtran}
\usepackage{amsfonts,amsmath,amssymb,amsthm,mathrsfs}
\usepackage[noadjust]{cite}
\usepackage{graphicx,multirow,bm}
\usepackage{color}
\usepackage{bm}
\makeatletter
\newtheoremstyle{mythm}{3pt}{3pt}{}{16pt}{\bfseries}{:}{.5em}{}
\theoremstyle{mythm}
\newtheorem{theorem}{Theorem}
\newtheorem{example}{Example}
\newtheorem{definition}{Definition}
\newtheorem{remark}{Remark}

\newtheorem{proposition}{Proposition}
\newtheorem{corollary}{Corollary}
\newtheorem{lemma}{Lemma}

\newtheorem{construction}{Construction}

\newcommand{\cC}{\mathcal{C}}

\newcommand{\cS}{\mathcal{S}}

\newcommand{\cV}{\mathcal{V}}
\newcommand{\cW}{\mathcal{W}}

\newcommand{\N}{\mathbb{N}}

\newcommand{\Z}{\mathbb{Z}}
\newcommand{\F}{\mathbb{F}}
\newcommand{\bS}{\mathbb{S}}

\DeclareMathOperator{\spn}{span}
\DeclareMathOperator{\colspn}{colspan}
\DeclareMathOperator{\rank}{rank}
\DeclareMathOperator{\diag}{diag}

\renewcommand{\leq}{\leqslant}

\renewcommand{\geq}{\geqslant}

\newcommand{\mathset}[1]{\left\{#1\right\}}
\newcommand{\abs}[1]{\left|#1\right|}

\newcommand{\floorenv}[1]{\left\lfloor #1 \right\rfloor}

\newcommand{\parenv}[1]{\left( #1 \right)}

\newcommand{\angleenv}[1]{\left\langle #1 \right\rangle}

\newcommand{\eqdef}{\triangleq}
\newcommand{\T}{\intercal}
\newcommand{\hS}{\widehat{S}}
\newcommand{\tS}{\widetilde{S}}

\newcommand{\hatt}{\widehat{t}}
\newcommand{\vv}[1]{\bm{#1}}
\newcommand{\tU}{\widetilde{U}}
\newcommand{\tW}{\widetilde{W}}
\newcommand{\tx}{\widetilde{x}}
\newcommand{\euler}[1]{\Z_{#1}^{\times}}

\begin{document}
\title{A Bound on the Minimal Field Size of LRCs, \\ and Cyclic
    MR Codes That Attain It
  }
 \author{
Han Cai,~\IEEEmembership{Member,~IEEE},
and Moshe Schwartz,~\IEEEmembership{Senior Member,~IEEE}
\thanks{The material in this paper was submitted in part to the IEEE International Symposium on Information Theory (ISIT 2022).}%
\thanks{Han Cai is with the School of Information Science and Technology,
  Southwest Jiaotong University,Chengdu, 610031, China
   (e-mail: hancai@aliyun.com).}%
\thanks{Moshe Schwartz is with the School
   of Electrical and Computer Engineering, Ben-Gurion University of the Negev,
   Beer Sheva 8410501, Israel
   (e-mail: schwartz@ee.bgu.ac.il).}%
\thanks{This research was supported in part by the German Research Foundation (DFG) with a German Israeli Project Cooperation (DIP) under grant no. PE2398/1-1.}
}
\date{}
\maketitle
\begin{abstract}
  We prove a new lower bound on the field size of locally repairable
  codes (LRCs). Additionally, we construct maximally recoverable (MR)
  codes which are cyclic. While a known construction for MR codes has
  the same parameters, it produces non-cyclic codes. Furthermore, we
  prove both necessary conditions and sufficient conditions that specify when the
  known non-cyclic MR codes may be permuted to become cyclic, thus
  proving our construction produces cyclic MR codes with new
  parameters. Furthermore, using our new bound on the field size, we
  show that the new cyclic MR codes have optimal field size in certain
  cases. Other known LRCs are also shown to have optimal field size in
  certain cases.
\end{abstract}


\section{Introduction}

\IEEEPARstart{I}{n} large-scale cloud storage and distributed file
systems, such as Amazon Elastic Block Store (EBS) and Google File
System (GoogleFS), disk failures are the norm and not the exception,
due to the sheer scale of the system.  To protect the data integrity,
coding theory is used to recover from data loss due to disk failures.
The simplest solution for those systems is a straightforward
replication of data packets across different disks.  However, this
solution is costly especially for large-scale systems since it suffers
from a large storage overhead.  As an alternative solution, erasure
codes such as $[n,k]$ maximum distance separable (MDS) codes, may be
employed as storage codes. These codes encode $k$ information symbols
to $n$ symbols and store them across $n$ disks, and they can recover
from the loss of any $n-k$ symbols. This scheme achieves a dramatic
improvement in redundancy compared with replication. However, for MDS
codes, even if one disk fails, the system needs to access $k$
surviving disks in order to recover the lost symbol, which makes the
repair process costly.

One method to improve the repair efficiently, suggested
in~\cite{gopalan2012locality}, is endow the code with a locality
property. This property allows a failed symbol to be recovered by
accessing only $r\ll k$ other symbols. Erasure codes with locality are
also called locally repairable codes (LRCs).  The original concept of
locality only works when exactly one erasure occurs (that is, one disk
fails). In the past decade, the notion of locality further generalized
in several directions.  For example, LRCs with
$(r,\delta)$-locality~\cite{prakash2012optimal} allow an erased symbol
to be recovered by reading $r$ other symbols, even if the repair set
suffered $\delta-1$ more erasures.  Other examples include: locality
which guarantees disjoint multiple repairable sets (also named as
availability)
\cite{wang2014repair,rawat2016locality,cai2018optimal,cai2019optimal},
locality which has a hierarchical
structure~\cite{sasidharan2015codes,chen2020cyclic}, and unequal
localities~\cite{kim2018locally,zeh2016bounds,hao2016constructions}.

Other code properties are also desirable. For a given code length $n$
and dimension $k$, we would like the Hamming distance to be as large
as possible, in order to maximize erasure-correcting
capabilities. Additionally, we would like the field size (or alphabet
size) to be as small as possible, in order to reduce the computation
complexity for coding and decoding. Other desirable properties may
include a cyclic structure for the code, since it allows for fast
encoding algorithms. Finally, even if the code has optimal distance,
we would like to be able to correct some pre-determined erasure
patterns beyond the minimum Hamming distance.

In the past a few years, many results have been obtained for
LRCs. Upper bounds on the minimum Hamming distance were proved, e.g.,
Singleton-type bounds
\cite{gopalan2012locality,prakash2012optimal,wang2015integer,cai2020optimal_bound},
and bounds related with the alphabet
size~\cite{cadambe2015bounds,agarwal2018combinatorial}. Optimal LRCs
(with respect to these bounds), were constructed, e.g.,
\cite{huang2013pyramid,rawat2013optimal,tamo2014family,song2014optimal,westerback2016combinatorics,
  chen2017constructions,martinez2019universal}.
In~\cite{guruswami2019long,cai2020optimal}, lower bounds on the field
size of optimal LRCs were derived for
$\delta=2$~\cite{guruswami2019long}, and $\delta\geq
2$~\cite{cai2020optimal}. Among the known optimal LRCs, some of them
also achieve order-optimal field
size~\cite{jin2019explicit,beemer2018explicit,xing2019construction,chen2020improved}
when $\delta=2$, and~\cite{cai2020optimal} when $\delta\geq
2$. Otherwise, constructions of optimal cyclic LRCs were introduced
in~\cite{tamo2016cyclic,chen2017constructions,qian2019new,chen2020cyclic,qiu2020new}.
When considering pre-determined recoverable erasure patterns beyond
the minimum Hamming distance, codes that can recover from all
information-theoretically recoverable erasure patterns are called
maximally recoverable (MR) codes~\cite{gopalan2012locality}, also
known as partial MDS codes~\cite{blaum2013partial}.
In~\cite{gopi2020maximally}, lower bounds on the field size
requirement for MR codes were introduced. For explicit constructions
of MR codes, the reader may refer
to~\cite{blaum2013partial,gopalan2014explicit,gabrys2018constructions,martinez2019universal,guruswami2020constructions,cai2022construction,gopi2020improved}. Notably,
there are MR codes have order-optimal field size (with respect to the
bound in~\cite{gopi2020maximally}): \cite{blaum2013partial} for a
single global parity check ($h=1$),
\cite{blaum2016construction,gopi2020maximally} for $h=2$,
\cite{gopalan2014explicit} for $h=3$ and $\delta=2$,
and~\cite{cai2022construction,gopi2020improved} for $h\leq \delta+1$ a
constant, and $n=\Theta(r^2)$.

The above summary shows how subsets of the mentioned desired
properties may be obtained simultaneously. However, to the best of our
knowledge, there are no explicit constructions that achieve all them,
namely, cyclic MR codes with optimal field size. In this paper, our motivation
is to construct cyclic MR codes with optimal field size, or order-optimal field size.
 To this end, we work both on constructions for cyclic MR codes, and a
 theoretic bound on the field size of optimal LRCs (containing
 MR codes as special cases).
In the first part of the paper we prove a
new general bound for optimal LRCs. We compare our new bound with the known bounds, and show that it is tighter for some parameters.
In the second part of the paper, we introduce a new construction for cyclic MR codes.
Our construction produces cyclic MR codes that share the same
parameters as one of the known non-cyclic constructions
in~\cite{gopi2020maximally}. We also show that under certain
conditions, the non-cyclic construction from~\cite{gopi2020maximally}
can be permuted to become a cyclic code, whereas in other cases it
cannot, thus proving our construction produces cyclic MR codes with
new parameters. As a byproduct of the proof, we characterize the algebraic structure of
repair sets for optimal cyclic LRCs, which results in strong new restrictions on
the parameters of optimal cyclic LRCs. Finally, we return to review
our bound on optimal LRCs,
and show that our construction has an
optimal field size when $r=2$. Since the bound is for general LRCs, as
a consequence we get that some known constructions have optimal field
size when $r=2$, a result which has not been claimed before.

The remainder of this paper is organized as follows. Section
\ref{sec-preliminaries} introduces some preliminaries about LRCs.
Section~\ref{sec-bound} proves a new bound on the field size of
LRCs. Section~\ref{sec-cyc-MR} describes a construction of cyclic MR
codes, as well as sufficient and necessary conditions under which a
known non-cyclic construction from~\cite{gopi2020maximally} may be
permuted to become cyclic. Section \ref{sec-conclusion} concludes this
paper with some remarks.

\section{Preliminaries}\label{sec-preliminaries}

In this section, we present notation and some necessary known results,
which are used throughout the paper. For a positive integer $n\in\N$,
we define $[n]=\mathset{0,1,\dots,n-1}$. If $m|n$ is a positive integer,
we denote
\[ \angleenv{m}\eqdef m\Z\cap [n] = \mathset{0,m,2m,\dots,n-m}.\]
Thus, $\angleenv{m}$ implicitly depends on $n$, whose value should be
understood from the context.

For any prime power $q$, let $\F_q$ denote the finite field of size
$q$, let $\F_q^m$ denote the set of vectors of length $m$ over $\F_q$,
and let $\F^{m\times n}_{q}$ denote the set of all possible $m\times
n$ matrices over $\F_q$.

An $[n,k]_q$ linear code, $\cC$, over $\F_q$, is a $k$-dimensional
subspace of $\F_q^n$. Such a code may be specified as the row-space of
a $k\times n$ generator matrix $G=(g_0,g_1,\dots,g_{n-1})$, where $g_i$
is a column vector of length $k$ for all $i\in [n]$. Specifically, it
is called an $[n,k,d]_q$ linear code if the minimum Hamming distance
of the code is $d$. For a subset $S\subseteq [n]$, we define
\begin{align*}
  \spn(S)&\eqdef \spn\mathset{g_i ~:~ i\in  S},\\
  \rank(S)&\eqdef \rank(\spn(S)).
\end{align*}
The code $\cC$ can also be specified by a parity-check matrix $H\in
\F^{(n-k)\times n}_{q}$, i.e., $\cC=\mathset{c\in
  \F^{n}_q~:~Hc^\T=0},$ where $\rank(H)=n-k$. Given a non-empty set of
coordinates, $S\subseteq[n]$, the punctured code $\cC|_S$ is the code
obtained from $\cC$ by deleting the code symbols at positions
$[n]\setminus S$. Thus, $\cC|_S$ is generated by $G|_S$ which is
obtained from $G$ by deleting the columns at $[n]\setminus S$.
Similarly, the shortened code $\cC|^S$ is the code whose parity matrix
is $H|_S$, namely, the matrix obtained from $H$ by deleting the
columns at $[n]\setminus S$.

An $[n,k]_q$ linear code, $\cC$, is said to be a cyclic code if
$c=(c_0,c_1,\cdots,c_{n-1})\in \cC$ implies that
$\sigma(c)\eqdef(c_{n-1},c_0,c_1,\cdots,$ $c_{n-2})\in \cC$, where
$\sigma$ is the cyclic shift operator by one place. It is well known
(see~\cite{MaSl77theory}) that a cyclic code with length $n$ over
$\F_q$ corresponds to a principal ideal of $\F_{q}[x]/(x^n-1)$. Thus,
let $\cC$ be generated by a monic polynomial $g(x)|(x^n-1)$, which is
called the generator polynomial of $\cC$. When $n|(q^m-1)$, assume
$\alpha$ is a primitive $n$th root of unity of $\F_{q^m}$, then the
cyclic code $\cC$ can be also be determined by the roots of $g(x)$,
i.e., $R_{\cC}=\mathset{\alpha^i ~:~ g(\alpha^i)=0}$.



We shall encounter many Vandermonde matrices in the following
section. Since we use a broader-than-usual definition for such
matrices, we give it here explicitly. Let
$\alpha_1,\dots,\alpha_n\in\F_q$ be $n$ distinct elements.  We say the
following $m\times n$ matrix is a Vandermonde matrix,
\[ \Pi \cdot D \cdot \begin{pmatrix}
  1 & 1 & \dots & 1 \\
  \alpha_0 & \alpha_1 & \dots & \alpha_{n-1} \\
  \vdots & \vdots & & \vdots \\
  \alpha_0^{m-1} & \alpha_1^{m-1} & \dots & \alpha_{n-1}^{m-1}
\end{pmatrix} D',\]
where $\Pi$ is a permutation matrix, and where $D$ and $D'$ are invertible
diagonal matrices. It is well known that the rank of such a matrix
is $\min\mathset{m,n}$.

\subsection{Locally Repairable Codes}

In~\cite{gopalan2012locality}, Gopalan \emph{et al.} introduced a
definition for the locality of code symbols.  For $j\in[n]$, the $j$th
code symbol, $c_j$, of an $[n, k,d]_q$ linear code, $\cC$, is said to
have locality $r$ if it can be recovered by accessing at most $r$
other symbols of $\cC$. This has been generalized
in~\cite{prakash2012optimal} to the following definition:

\begin{definition}\label{def_r_delta_i}
  Let $\cC$ be an $[n,k,d]_q$ linear code, and let $G$ be a generator
  matrix for it. For $j\in[n]$, the $j$th code symbol, $c_j$, of
  $\cC$, is said to have $(r, \delta)$-locality if there exists a
  subset $S_j\subseteq [n]$ such that:
  \begin{itemize}
  \item $j\in S_j$ and $|S_j|\leq r+\delta-1$; and
  \item the minimum Hamming distance of the punctured code
    $\cC|_{S_j}$ is at least $\delta$.
  \end{itemize}
  In that case, the set $S_j$ is also called a repair set of $c_j$.
  The code $\cC$ is said to have information $(r,\delta)$-locality if
  there exists $S\subseteq [n]$ with $\rank(S)=k$ such that for each
  $j\in S$, $c_j$ has $(r, \delta)$-locality. Furthermore, the code
  $\cC$ is said to have all-symbol $(r,\delta)$-locality if all the
  code symbols have $(r,\delta)$-locality.
\end{definition}

Thus, the definition of symbol locality
from~\cite{gopalan2012locality} is the special case of $\delta=2$ in
the definition
from~\cite{prakash2012optimal}. In~\cite{prakash2012optimal} (and for
the case $\delta=2$, originally~\cite{gopalan2012locality}), the
following upper bound on the minimum Hamming distance of linear codes
with information $(r,\delta)$-locality is derived.
\begin{theorem}[\cite{prakash2012optimal}] \label{lemma_bound_i}
  For an $[n,k,d]_q$ linear code with information $(r,\delta)$-locality,
\[ d\leq n-k+1-\left(\left\lceil\frac{k}{r}\right\rceil-1\right)(\delta-1).\]
\end{theorem}

Codes with information $(r,\delta)$-locality are said to be
\emph{optimal locally repairable codes (optimal LRCs)} if their
minimum Hamming distance attains the bound of
Theorem~\ref{lemma_bound_i} with equality. It is known that optimal LRCs
with all-symbol $(r,\delta)$-locality have a specific structure to
their repair sets.

\begin{theorem}[\cite{song2014optimal,cai2020optimal}]\label{lemma_repair_sets}
  Let $\cC$ be an optimal $[n,k,d]_q$ LRC with all-symbol
  $(r,\delta)$-locality. Let $\Gamma\subseteq 2^{[n]}$ be the set of
  all possible repair sets. Write $k=ru+v$, for integers $u$ and $v$,
  and $0\leq v\leq r-1$.  If $(r+\delta-1) | n$, $k>r$, and
  additionally, $u\geq 2(r-v+1)$ or $v=0$, then there exists a subset $\cS\subseteq \Gamma$, such that:
  \begin{itemize}
  \item All $S\in\cS$ are of cardinality $\abs{S}=r+\delta-1$, and $\cS$ is a partition of $[n]$.
  \item
    For any $S\in\cS$, $\cC|_{S}$ is an $[r+\delta-1,r,\delta]_q$ MDS
    code.
  \end{itemize}
\end{theorem}

\begin{remark}
  The partitioning of $[n]$ by repair sets was first proved
  in~\cite{song2014optimal} only for the case $r|k$, i.e., $v=0$.
  Recently, this property was proved in~\cite{cai2020optimal} also for
  the case $u\geq 2(r-v+1)$.
\end{remark}

In~\cite{guruswami2019long}, Guruswami \emph{et al.} asked a
fundamental interesting question: How long can an optimal LRC with
$(r,\delta=2)$-locality be? They derived the following upper bound on
the code length.

\begin{theorem}[\cite{guruswami2019long}]\label{lemma_bound_2}
Let $\cC$ be an optimal $[n,k,d]_q$ LRC with all-symbol
$(r,2)$-locality. If $d\geq 5$, $k>r$, $(r+1)|n$, and additionally,
$r|k$ or $k\geq 2r^2+2r-(2r-1)(k \bmod r)$, then
\begin{equation}
n=
\begin{cases}
O\parenv{dq^{\frac{4(d-2)}{d-a}-1}},& \text{if }a=1,2,\\
O\parenv{dq^{\frac{4(d-3)}{d-a}-1}},& \text{if }a=3,4,\\
\end{cases}
\end{equation}
where $a\in\mathset{1,2,3,4}$, and $a\equiv d\pmod{4}$.
\end{theorem}

In~\cite{cai2020optimal}, this problem is further considered for
optimal LRCs with all-symbol $(r,\delta)$-locality, $\delta\geq 2$.

\begin{theorem}[\cite{cai2020optimal}]\label{lemma_bound_delta>2}
Let $n=w(r+\delta-1)$, $\delta>2$, $k=ur+v$, $0\leq v\leq r-1$, and
additionally, $u\geq 2(r-v+1)$ or $v=0$, where all parameters are
integers. Assume that there exists an optimal $[n,k,d]_q$ linear code
$\cC$ with all-symbol $(r,\delta)$-locality, and define
$t=\floorenv{(d-1)/\delta}$.  If $t\geq 2$, then
\begin{align*}
n&\leq
\begin{cases}
\frac{(t-1)(r+\delta-1)}{2r(q-1)}q^{\frac{2(w-u)r-2v}{t-1}} & \text{ if } t \text{ is odd}  \\
\frac{t(r+\delta-1)}{2r(q-1)}q^{\frac{2(w-u)r-2v}{t}} &\text{ if } t \text{ is even}  \\
\end{cases}\\
&=O\parenv{\frac{t(r+\delta)}{r}q^{\frac{(w-u)r-v}{\floorenv{t/2}}-1}},
\end{align*}
where $w-u$ can also be rewritten as $w-u=\lfloor(d-1+v)/(r+\delta-1)\rfloor$.
\end{theorem}

\subsection{Maximally Recoverable Codes}

Maximally recoverable (MR) codes are an extremal case of LRCs, that
maximize the erasure-repair capability.

\begin{definition}
  \label{def:mrcodes}
  Let $\cC$ be an $[n,k,d]_q$ code with all-symbol
  $(r,\delta)$-locality, and define $\cS\eqdef\{S_i ~:~ i\in [n]\}$,
  where $S_i$ is a repair set for coordinate $i$. The code $\cC$ is
  said to be a \emph{maximally recoverable (MR) code} if $\cS$ is a
  partition of $[n]$, and for any $R_i\subseteq S_i$ such that
  $|S_i\setminus R_i|=\delta-1$, the punctured code
  $\cC|_{\cup_{i\in[n]}R_i}$ is an MDS code.
\end{definition}

In general, $S_i$ for $i\in[n]$, are not required to be of the same
size. However, from an application point of view, equal-sized repair sets
simplify the implementation, bringing us to the following definition:

\begin{definition}\label{def_MRC}
  Let $\cC$ be an $[n,k,d]_q$ MR code, as in
  Definition~\ref{def:mrcodes}.  If each $S_i\in\cS$ is of size
  $|S_i|=r+\delta-1$ (implying $r+\delta-1|n$), we define
  \begin{eqnarray*}
    m\eqdef \frac{n}{r+\delta-1}, \qquad h\eqdef mr-k.
  \end{eqnarray*}
  Then $\cC$ is said to be an $(n,r,h,\delta,q)$-MR code.
\end{definition}

We first note that it is easy to verify that $(n,r,h,\delta,q)$-MR
codes are optimal $[n,k,d]_q$ LRCs with all-symbol
$(r,\delta)$-locality. We can regard each codeword of an
$(n,r,h,\delta,q)$-MR code, as an $m\times(r+\delta-1)$ array, by
placing each repair set in $\cS$ as a row, when $\cS$ forms a partition
of $[n]$. In this way,
$(n,r,h,\delta,q)$-MR codes match the definition of partial MDS (PMDS)
codes, as defined in \cite{blaum2013partial}. When implemented in a
distributed-storage setting, each entry of a codeword array
corresponds to a sector, each column of the array corresponds to a
disk, and each row to a stripe. Thus, an $(n,r,h,\delta,q)$-MR code
can recover from $\delta-1$ sector erasures in each stripe, and
additional $h$ erased sectors anywhere. We mention in passing that a
more restricted type of codes, called sector-disk (SD) codes, are
capable of recovering from $\delta-1$ disk erasures, and additional
$h$ erased sectors (see~\cite{plank2014sector,cai2020optimal_GPMDS}).


Paralleling the general case of optimal LRCs, it is interesting to ask
what is the minimum alphabet size required by MR codes.

\begin{lemma}[{\cite[Theorem I.1]{gopi2020maximally}}]
  \label{lemma_lower_bound_F}
  Let $\cC$ be an $(n,r,h,\delta,q)$-MR code, $h\geq 2$. If
  $m\eqdef\frac{n}{r+\delta-1}\geq 2$, then
  \begin{equation*}
    q= \Omega(n r^{\varepsilon}),
  \end{equation*}
  where
  $\varepsilon=\min\{\delta-1,h-2\lceil\frac{h}{m}\rceil\}/\lceil\frac{h}{m}\rceil$,
  and where $h$ and $\delta$ are regarded as constants. Additionally,
  \begin{enumerate}
  \item If $m\geq h$:
    \[
    q= \Omega\parenv{n r^{\min\{\delta-1,h-2\}}}.
    \]
  \item If $m\leq h$, $m|h$, and $\delta-1\leq h-\frac{2h}{m}$:
    \[q= \Omega\parenv{n^{1+\frac{m(\delta-1)}{h}}}.\]
  \item If $m\leq h$, $m|h$, and $\delta-1> h-\frac{2h}{m}$:
    \[q= \Omega\parenv{n^{m-1}}.\]
  \end{enumerate}
\end{lemma}

\begin{remark}
For the case $h=1$, the field size requirement of an
$(n,r,h,\delta,q)$-MR code may be as small as $q=\Theta(r+\delta-1)$.
This is attainable since the punctured code over any repair set
together with the single global parity check is an
$[r+\delta,r,\delta+1]_q$ MDS code when $(r+\delta-1)|k$ or $u\geq
2(r-v+1)$, where $k=ur+v$ with $0\leq v\leq r-1$
(see~\cite{cai2020optimal}).
\end{remark}

\begin{definition}\label{def_ord_opt}
A family of $(n,r,h,\delta,q)$-MR codes has \emph{order-optimal
  field size} if it attains one of the bounds of
Lemma~\ref{lemma_lower_bound_F} asymptotically for $h\geq 2$, or if it
has $q=\Theta(r+\delta-1)$ for $h=1$.
\end{definition}

\section{A New Bound on Optimal LRCs}\label{sec-bound}

In this section we present a new bound on the parameters of optimal
LRCs with all-symbol $(r,\delta)$-locality. To that end, we first prove
bounds for optimal LRCs with small minimum Hamming distance, distinguishing between the two cases of $2\mid r$
and $2\nmid r$. The proof strategy of both bounds is showing that the existence of such codes forces the existence of many subspaces, any two of which intersect only trivially. We then recall a parameter-reduction
lemma, which reduces optimal LRCs with a large minimum Hamming distance into optimal LRCs with a smaller
one. Combining these together results in the main bound.
We note that the new bound is not
specific to MR codes or to cyclic codes, but instead applies to optimal LRCs. The bound does, however, require
certain divisibility conditions, which are common to several
constructions of optimal LRCs, among them, MR codes.

\begin{lemma}\label{lemma_bound_even}
Let $\cC$ be an optimal $[n=(u+1)(r+\delta-1),ur,r+2\delta-1]_q$ LRC
with all-symbol $(r,\delta)$-locality. If $2|r$, then
\[
u+1\leq (q^{r/2}+1)\Big/\floorenv{\frac{2r+2\delta-2}{r}}.
\]
\end{lemma}

\begin{IEEEproof}
Denote $t\eqdef\lceil(2r+2\delta-2)/r\rceil$ and
$t'\eqdef\lfloor(2r+2\delta-2)/r\rfloor$. By
Theorem~\ref{lemma_repair_sets}, the code $\cC$ has a parity-check
matrix of the following form,
\[P=\begin{pmatrix}
V_{0,0}&V_{0,1}&\cdots&V_{0,t-1}&0&0&\cdots&0&\cdots&0&0&\cdots&0\\
0&0&\cdots &0&V_{1,0}&V_{1,1}&\cdots&V_{1,t-1}&\cdots&0&0&\cdots&0\\
\vdots&\vdots& &\vdots&\vdots&\vdots& &\vdots& \ddots &\vdots&\vdots& &\vdots\\
0&0&\cdots&0&0&0&\cdots&0&\cdots&V_{u,0}&V_{u,1}&\cdots&V_{u,t-1}\\
W_{0,0}&W_{0,1}&\cdots&W_{0,t-1}&W_{1,0}&W_{1,1}&\cdots&W_{1,t-1}&\cdots&W_{u,0}&W_{u,1}&\cdots&W_{u,t-1}\\
\end{pmatrix},\]
where $V_{i,j}\in \F^{(\delta-1)\times (r/2)}_{q}$, $W_{i,j}\in
\F^{r\times (r/2)}_{q}$ for $i\in [u+1]$ and $j\in [t-1]$, and
$(V_{i,0}\,V_{i,1},\cdots,V_{i,t-1})$ is parity-check matrix of an
$[r+\delta-1,r,\delta]_q$ MDS code for $i\in [u+1]$. This implies that
the matrices $V_{i,t-1}$ and $W_{i,t-1}$, $i\in[u+1]$, have $\frac{r}{2}$
columns each when $r|(2r+2\delta-2)$, and
$(2r+2\delta-2)\bmod\frac{r}{2}$ otherwise.

Let us consider the following square
$(r+2\delta-2)\times(r+2\delta-2)$ matrices,
\[
E_{a,b,i,j}\eqdef\begin{pmatrix}
V_{a,0}&\cdots&V_{a,i-1}&V_{a,i+1}&\cdots &V_{a,t-1}&0&\cdots&0&0&\cdots&0\\
0&\cdots&0&0&\cdots&0&V_{b,0},&\cdots&V_{b,j-1}&V_{b,j+1}&\cdots &V_{b,t-1}\\
W_{a,0},&\cdots&W_{a,i-1}&W_{a,i+1}&\cdots &W_{a,t-1}&W_{b,0},&\cdots&W_{b,j-1}&W_{b,j+1}&\cdots &W_{b,t-1}
\end{pmatrix},
\]
where $a,b\in [u+1]$, $a\neq b$, and $i,j\in [t']$. Since the minimum
Hamming distance of $\cC$ is $r+2\delta-1$, any $r+2(\delta-1)$
columns from $P$ are linearly independent. This implies that the
matrices $E_{a,b,i,j}$ defined above have full rank.

Recall that $(V_{a,0},V_{a,1},\cdots,V_{a,t-1})$ is a parity-check
matrix of an $[r+\delta-1,r,\delta]_q$ MDS code. Thus,
$(V_{a,1},\cdots,V_{a,i-1},V_{a,i+1},$ $\cdots,V_{a,t-1})$ is an invertible
$(\delta-1)\times(\delta-1)$ matrix. A similar claim follows for
$(V_{b,1},\cdots,V_{b,j-1},V_{b,j+1},\cdots,V_{b,t-1})$. Hence, after
simple column and row operations, the full rank of $E_{a,b,i,j}$
implies that
\[
\begin{pmatrix}
0&V_{a,1}&\cdots&V_{a,i-1}&V_{a,i+1}&\cdots &V_{a,t-1}&0&0&\cdots&0&0&\cdots&0\\
0&0&\cdots&0&0&\cdots&0&0&V_{b,1}&\cdots&V_{b,j-1}&V_{b,j+1}&\cdots &V_{b,t-1}\\
W^*_{a,i}&0&\cdots&0&0&\cdots &0&W^*_{b,j}&0&\cdots&0&0&\cdots &0\\
\end{pmatrix}
\]
has full rank, implying also that
\begin{equation}\label{eqn_rank_Eab}
\rank(W^*_{a,i},W^*_{b,j})=r,
\end{equation}
for $a,b\in [u+1]$, $a\neq b$, and $i,j\in [t']$. We also mention that
if either $i=0$ or $j=0$, natural adjustments need to be made, that
is, zeroing $V_{a,1}$ instead of $V_{a,0}$, and $V_{b,1}$ instead of
$V_{b,0}$.

Next, assume $a\in[u+1]$, and $i,j\in[t']$, $i\neq j$. We pick only
$r+\delta-1$ columns from $P$, which must therefore be linearly
independent, giving us,
\begin{equation}
  \label{eq:sameab}
\begin{split}
r+\delta-1&=\rank\begin{pmatrix}
V_{a,0}&V_{a,1}&\cdots&V_{a,t-1}\\
W_{a,0}&W_{a,1}&\cdots&W_{a,t-1}\\
\end{pmatrix}\\
&=\rank\begin{pmatrix}
V_{a,0}&V_{a,0}&V_{a,1}&\cdots&V_{a,t-1}\\
W_{a,0}&W_{a,0}&W_{a,1}&\cdots&W_{a,t-1}\\
\end{pmatrix}\\
&=\rank\begin{pmatrix}
V_{a,0}&0&V_{a,1}&\cdots&V_{a,t-1}\\
W_{a,0}&W^*_{a,j}&W_{a,1}&\cdots&W_{a,t-1}\\
\end{pmatrix}\\
&=\rank\begin{pmatrix}
V_{a,0}&0&V_{a,1}&\cdots&V_{a,i-1}&V_{a,i+1}&\cdots&V_{a,t-1}\\
W_{a,0}&W^*_{a,j}&W_{a,1}&\cdots&W_{a,i-1}&W_{a,i+1}&\cdots&W_{a,t-1}\\
\end{pmatrix}\\
&=\rank\begin{pmatrix}
0&0&V_{a,1}&\cdots&V_{a,i-1}&V_{a,i+1}&\cdots&V_{a,t-1}\\
W^*_{a,i}&W^*_{a,j}&0&\cdots&0&0&\cdots&0\\
\end{pmatrix}.
\end{split}
\end{equation}
We explain why the fourth equality holds in more detail. The column
operations performed in order to obtain $W^*_{a,j}$ may be written as
\[
\begin{pmatrix}
0\\
W^*_{a,j}\\
\end{pmatrix}=\sum_{\tau\in[t]\setminus\mathset{j}}\begin{pmatrix}
V_{a,\tau}\\
W_{a,\tau}\\
\end{pmatrix}E_\tau,
\]
where $E_0=I$ is the identity matrix. It then follows that
\[
(V_{a,0},\dots,V_{a,j-1},V_{a,j+1},\dots,V_{a,t-1})\cdot
\begin{pmatrix}
  E_0 \\
  \vdots\\
  E_{j-1} \\
  E_{j+1} \\
  \vdots \\
  E_{t-1}
\end{pmatrix} = 0.
\]
Since $(V_{a,0},\dots,V_{a,j-1},V_{a,j+1},\dots,V_{a,t-1})$ is a
parity-check matrix for an
$[\frac{r}{2}+\delta-1,\frac{r}{2},\delta]_q$ MDS code, we have that
the matrix $(E_0^\T,\dots,E_{j-1}^\T,E_j^\T,\dots,E_{t-1}^\T)$ is a
generator matrix for that code. Hence, any $\frac{r}{2}$ columns of it
are linearly independent. In particular, that means $E_i$ is
invertible.  We can therefore write,
\begin{equation*}
\begin{pmatrix}
V_{a,i}\\
W_{a,i}\\
\end{pmatrix}=-\sum_{\tau\in[t]\setminus\mathset{i,j}}\begin{pmatrix}
V_{a,\tau}\\
W_{a,\tau}\\
\end{pmatrix}E_{\tau}E^{-1}_i+\begin{pmatrix}
0\\
W^*_{a,j}\\
\end{pmatrix}E^{-1}_i.
\end{equation*}
This completes the detailed explanation for the fourth equality
in~\eqref{eq:sameab}. The main observation is that~\eqref{eq:sameab}
gives
\begin{equation}\label{eqn_rank_Eaa}
\rank(W^*_{a,i},W^*_{a,j})=r,
\end{equation}
for $a\in [u+1]$, and $i,j\in [t']$, $i\neq j$. Again, if $i=0$ or
$j=0$, a natural adjustment needs to be made.

Let us define the following set of subspaces
\[ \cW\eqdef \mathset{ \colspn(W^*_{a,i}) ~:~ a\in[u+1], i\in[t'] },\]
where $\colspn(\cdot)$ of a matrix denotes its column
space. By~\eqref{eqn_rank_Eab} and~\eqref{eqn_rank_Eaa} we learn that
$\cW$ contains only $\frac{r}{2}$-dimensional spaces, which are all distinct,
hence
\[ \abs{\cW}=(u+1)t' = (u+1)\floorenv{\frac{2r+2\delta-2}{r}}.\]
Additionally, any two subspaces from $\cW$ intersect only trivially,
hence
\[(u+1)\floorenv{\frac{2r+2\delta-2}{r}}=\abs{\cW}\leq \frac{q^r-1}{q^{r/2}-1}=q^{r/2}+1.\]
Rearranging this gives the desired claim.
\end{IEEEproof}

For the case $2\nmid r$, we also have a similar lemma.

\begin{lemma}\label{lemma_bound_odd}
Let $\cC$ be an optimal $[n=(u+2)(r+\delta-1),ur,2r+3\delta-2]_q$ LRC
with all-symbol $(r,\delta)$-locality. If $2\nmid r$, then
\[
u\leq q^{(r+1)/2}.
\]
\end{lemma}

\begin{IEEEproof}
By Theorem~\ref{lemma_repair_sets}, and after simple row operations, the
code $\cC$ has a parity-check matrix of the following form,
\[
P=\begin{pmatrix}
I_{\delta-1}&V_{0,0}&V_{0,1}&0&0&0&\cdots&0&0&0\\
0&0&0&I_{\delta-1}&V_{1,0}&V_{1,1}&\cdots&0&0&0\\
\vdots&\vdots&\vdots&\vdots&\vdots&\vdots&\ddots&\vdots&\vdots&\vdots\\
0&0&0&0&0&0&\cdots&I_{\delta-1}&V_{u+1,0}&V_{u+1,1}\\
0&W_{0,0}&W_{0,1}&0&W_{1,0}&W_{1,1}&\cdots&0&W_{u+1,0}&W_{u+1,1}\\
\end{pmatrix},\]
where $I_{\delta-1}$ is the $(\delta-1)\times(\delta-1)$ identity
matrix, $V_{i,1}\in \F^{(\delta-1)\times ((r-1)/2)}_{q}$, $V_{i,2}\in
\F^{(\delta-1)\times ((r+1)/2)}_{q}$, $W_{i,1}\in \F^{2r\times
  ((r-1)/2)}_{q}$, $W_{i,2}\in \F^{2r\times ((r+1)/2)}_{q}$, and
$(I_{\delta-1}, V_{i,1},V_{i,2})$ is a parity-check matrix of an
$[r+\delta-1,r,\delta]_q$ MDS code, for all $i\in [u+2]$.

Consider the following square $(2r+3\delta-3)\times(2r+3\delta-3)$ matrices,
\[
E_{a,b}\eqdef\begin{pmatrix}
I_{\delta-1}&V_{0,0}&V'_{0,1}&0&0&0&0\\
0&0&0&I_{\delta-1}&V_{a,1}&0&0\\
0&0&0&0&0&I_{\delta-1}&V_{b,1}\\
0&W_{0,0}&W'_{0,1}&0&W_{a,1}&0&W_{b,1}\\
\end{pmatrix},
\]
where $a,b\in [u+2]\setminus\mathset{0}$, $a\neq b$, and where
$V'_{0,1}$ and $W'_{0,1}$ are the first $\frac{r-1}{2}$ columns of
$V_{0,1}$ and $W_{0,1}$, respectively. Since the minimum Hamming
distance of $\cC$ is $2r+3\delta-2$, any $2r+3\delta-3$ columns from
$P$ are linearly independent, and in particular,
\[ \rank(E_{a,b})=2r+3\delta-3.\]
This implies that
\[ \rank( W_{0,0}, W'_{0,1}, W_{a,1}, W_{b,1})=2r.\]
By the size of the matrices, we also must have
\[ \rank( W_{a,1},W_{b,1})=r+1,\]
and also
\[\rank(W_{a,1})=\rank(W_{b,1})=\frac{r+1}{2}.\]

We now denote $U'=\colspn(W_{0,0},W'_{0,1})\subseteq
\F_q^{2r}$. Obviously, $\dim(U')=r-1$. Let us arbitrarily choose an
$(r+1)$-dimensional subspace $\tU\subseteq\F_q^{2r}$ such that
$\F_q^{2r}=U'+\tU$, namely, $\dim(\tU)=r+1$ and $U'\cap
\tU=\mathset{0}$.  For any vector $x\in\F_q^{2r}$, let
$\tx\in\F_q^{2r}$ denotes its projection onto $\tU$, that is,
$\tx\in\tU$ is the unique vector such that $x=x'+\tx$, with $x'\in
U'$. For any $a\in[u+2]\setminus\mathset{0}$, we then construct
$\tW_{a,1}$ from $W_{a,1}$ by replacing each column vector with its
projection onto $\tU$. It then follows, that for all
$a,b\in[u+2]\setminus\mathset{0}$, $a\neq b$,
\[ \rank(W_{0,0},W'_{0,1},\tW_{a,1},\tW_{b,1})=2r,\]
and also
\[ \rank(\tW_{a,1})=\rank(\tW_{b,1})=\frac{r+1}{2}.\]

Let us construct the set of subspaces
\[ \cW\eqdef \mathset{\colspn(\tW_{a,1}) ~:~ a\in[u+2]\setminus\mathset{0}}.\]
By the previous discussion, $\cW$ contains $u+1$ subspaces of $\tU$,
each of dimension $\frac{r+1}{2}$, any two of which intersect only
trivially.  Additionally, the sum of any two subspaces from $\cW$,
summed together with the fixed $(r-1)$-dimensional subspace
$\colspn(W_{0,0})+\colspn(W'_{0,1})$, gives $\F_q^{2r}$.  Thus,
\[ u+1=\abs{\cW} \leq \frac{q^{r+1}-1}{q^{(r+1)/2}-1} = q^{(r+1)/2}+1,\]
completing the proof.
\end{IEEEproof}

The final component in our main bounding theorem is a
parameter-reduction lemma.  This lemma was proved
in~\cite{cai2020optimal}.

\begin{lemma}[\cite{cai2020optimal} Corollary 2]\label{lemma_reduce_dis}
Let $n=m(r+\delta-1)$, $\delta\geq 2$, $k=ur+v>r$, and additionally,
$r|k$ or $u\geq 2(r+1-v)$, where all parameters are integers. If there exists an optimal $[n,k,d]_q$
linear code $\cC$ with $d>r+\delta$ and all-symbol $(r,\delta)$-locality,
then there exists an optimal linear code $\cC'$ with all-symbol
$(r,\delta)$-locality and parameters $[n-\epsilon(r+\delta-1),k,d'=d-\epsilon(r+\delta-1)]_q,$
where $\epsilon\leq \lceil (d-1)/(r+\delta-1)\rceil-1$.
\end{lemma}

Let us now state and prove our main bound. The next theorem gives a
lower bound on the size of the field required for LRCs with all-symbol
$(r,\delta)$-locality.

\begin{theorem}\label{thm_bound}
Let $\cC$ be an optimal $[n,k,d]_q$ linear code with all-symbol
$(r,\delta)$-locality. Assume $n=m(r+\delta-1)$, $k=ur$, $u\geq 2$.
If $2|r$ and $m\geq u+1$ then,
\[
q \geq
  \psi\parenv{\parenv{\parenv{\frac{k}{r}+1}\floorenv{\frac{2r+2\delta-2}{r}}-1}^{\frac{2}{r}}},
\]
where $\psi(x)$ is the smallest prime power greater or equal to $x$.
If $2\nmid r$ and $m\geq u+2$ then
\[ q \geq \psi\parenv{\parenv{\frac{k}{r}}^{\frac{2}{r+1}}}.\]
\end{theorem}

\begin{IEEEproof}
  The proof is straightforward. Apply Lemma~\ref{lemma_reduce_dis}
  until reaching the required conditions of either
  Lemma~\ref{lemma_bound_even} or Lemma~\ref{lemma_bound_odd}, and
  then use them.
\end{IEEEproof}

\begin{remark}
Assume that there exists an optimal $[n,k,d]_q$ LRC with all-symbol
$(r,\delta)$-locality, and define $t=\floorenv{(d-1)/\delta}\geq
2$. Rewriting the bound of Theorem~\ref{lemma_bound_delta>2} (which we cite from~\cite{cai2020optimal}) in a slightly looser yet more convenient way,
\begin{equation*}
q=\Omega\left(\left(\frac{nr}{t(r+\delta)}\right)^{\frac{\floorenv{(d-1)/\delta}}{2\floorenv{\frac{d-1+v}{r+\delta-1}} r-2v}}\right).
\end{equation*}
Thus, when only $k$ and $n$ tend to infinity, and the other parameters are constant, if
\[{\frac{\floorenv{(d-1)/\delta}}{2\floorenv{\frac{d-1+v}{r+\delta-1}} r-2v}}<\frac{2}{r+1},\]
then the exponent in the bound of Theorem~\ref{thm_bound} is higher, and it may outperform the known bound of Theorem~\ref{lemma_bound_delta>2} (where we denote $k=ur+v$ and $0\leq v\leq r-1$).
\end{remark}

Having seen that the new bound of Theorem~\ref{thm_bound} may provide an improvement, we focus on a single case. More specifically, the case of $r=2$ is of particular interest, since we
can then use Theorem~\ref{thm_bound} to prove that some known LRCs
have optimal field size.

We first consider some Tamo-Barg
codes~\cite{tamo2014family}.

\begin{lemma}[\cite{tamo2014family}]\label{lemma_TB_code}
Let $q$ be a prime power, $q=r+\delta-1$, then there exists an optimal
LRC with all-symbol $(r,\delta)$-locality and parameters
$[q^b,ur,(q^{b-1}-u)q+\delta]_{q^b}$, where $b\geq 2$ and
$0<u<q^{b-1}$.
\end{lemma}

\begin{corollary}
  Let $\cC$ be a code from Lemma~\ref{lemma_TB_code} with $r=2$ and
  $u=q^{b-1}-1$. If $q^b-1$ is not a prime power then $\cC$ has
  optimal field size.
\end{corollary}

\begin{example}
Let $n=2^{4}$, $r=2$, $\delta=3$, then by Lemma~\ref{lemma_TB_code}
there exists an optimal LRC with all-symbol $(2,3)$-locality and
parameters $[16,6,7]_{2^4}$, which has optimal field size since $15$
is not a prime power.
\end{example}

We now examine a construction of cyclic optimal LRCs
from~\cite{tamo2016cyclic}.

\begin{lemma}[\cite{tamo2016cyclic}]\label{lemma_LRC_tamo16}
Let $r=2$, $n=m(r+\delta-1)=q^b-1$, and $k=ur+v$ with $0\leq v<r$,
where $q^b$ is prime power.  Then there exists a cyclic optimal LRC
with all-symbol $(2,\delta)$-locality and parameters
$[q^b-1,k,d]_{q^b}$.
\end{lemma}

\begin{corollary}
  Let $\cC$ be a code from Lemma~\ref{lemma_LRC_tamo16} with $m=u+1$
  and $v=0$.  If neither $q^b-2$, nor $q^b-1$, are prime powers, then
  $\cC$ has optimal field size.
\end{corollary}

\begin{example}
  Let $n=2^{6}-1$, $r=2$, and $\delta=2$. Then by
  Lemma~\ref{lemma_LRC_tamo16}, there exists a cyclic optimal LRC with
  all-symbol $(2,2)$-locality, and parameters $[63,40,5]_{2^6}$, which
  has optimal field size since both $62$ and $63$ are not prime
  powers.
\end{example}

Yet another construction of cyclic optimal LRCs comes
from~\cite{chen2017constructions}.

\begin{lemma}[\cite{chen2017constructions}]\label{lemma_CLRC_Chen}
  Let $r=2$, $\delta=2$, $n=m(r+\delta-1)=3m=q^b+1$, and $k=2u$, with
  $u$ an even integer, and where $q^b$ is prime power.  Then there
  exists a cyclic optimal LRC with all-symbol $(2,2)$-locality and
  parameters $[q^b+1=3m,2u,d]_{q^m}$.
\end{lemma}

\begin{corollary}\label{corollary_exa_opt}
  Let $\cC$ be a code from Lemma~\ref{lemma_CLRC_Chen} with $m=u+1$.
  Then $\cC$ has optimal field size.
\end{corollary}

\begin{example}\cite{chen2017constructions}
Let $n=9=2^3+1$, $r=2$, $\delta=2$, $k=4$, then there exists a cyclic optimal
$[9,4,5]_8$-LRC, which has optimal field size.
\end{example}

Note that an $(n,r=2,h=2,\delta,q)$-MR code
is also an $[n=(u+1)(\delta+1),2u,2\delta+1]_q$ optimal LRC.
The following corollary can be derived directly from Lemma~\ref{lemma_bound_even}.

\begin{corollary}\label{Lemma_length_r=2}
  Let $\cC$ be an $(n,r=2,h=2,\delta,q)$-MR code. Then $q\geq n-1$.
\end{corollary}


\section{Cyclic Maximally Recoverable (MR) Codes}
\label{sec-cyc-MR}

We divide this section into two parts. In the first part we construct
cyclic MR codes, and show that for certain parameters they have the
exact optimal field size. The main idea behind our construction is to carefully choose the roots for the cyclic code we construct, in a way that produces an MR code.
In the second part we study a known class of
MR codes which are non-cyclic, but have the same parameters as the
cyclic codes we construct. We then show that these non-cyclic codes
can sometimes be permuted to obtain cyclic MR codes. For this part, as tools to
prove the main results, we characterize the algebraic structure of
the repair sets of \emph{cyclic} optimal LRCs (Theorem~\ref{theorem_repair_set_cyc}) and the structure of punctured codes and shortened
codes over repair sets (Corollary~\ref{corollary_ parity}). On the one hand, we prove that for some parameters
the known construction can be permuted to obtain cyclic MR codes by finding suitable
permutations (Theorem~\ref{th:multi_permu}). On the other hand, we show that such a permutation is not always possible (Theorem \ref{th:cyclic}). By combining the two parts, we obtain our main result.

\subsection{A New Construction}

We immediately present our construction for cyclic MR codes. It is
inspired by the construction of~\cite{tamo2016cyclic}.

\begin{construction}\label{cons_CPMDS}
  Let $b,r,\delta\geq 2$ be integers, $q$ a prime power, $n=q^b-1$,
  $\alpha\in\F_{q^b}$ a primitive element, $a=(r+\delta-1)|(q-1)$, and
  $m=n/a$ such that $\gcd(\delta,m)=1$.  Define
  \[ R\eqdef\mathset{\alpha^{ja+t}~:~1\leq j\leq m,1\leq t\leq \delta-1}\cup\mathset{1,\alpha^{\delta}}.
  \]
  The constructed code, $\cC$, is the cyclic code of length $n$ over
  $\F_{q^b}$ with root set $R$.
\end{construction}

Our goal is now to show that the code from
Construction~\ref{cons_CPMDS} is indeed a cyclic MR code. However, in
order to do so we require a technical lemma.

\begin{lemma}\label{lemma_full_rank}
  Assume the setting and notation of Construction~\ref{cons_CPMDS}.
  Denote $\beta=\alpha^{m}$, and $\gamma=\alpha^\delta$. Assume
  $T_i=\mathset{t_{i,1},\dots,t_{i,\delta}}\subseteq[a]$ for $i=1,2$. Then
  for any $i_1,i_2\in[m]$, $i_1\neq i_2$, the matrix
\begin{equation*}
M=\parenv{\begin{array}{cccc|cccc}
\beta^{t_{1,1}}&\beta^{t_{1,2}}&\cdots&\beta^{t_{1,\delta}}&0&0&\cdots&0\\
\beta^{2t_{1,1}}&\beta^{2t_{1,2}}&\cdots&\beta^{2t_{1,\delta}}&0&0&\cdots&0\\
\vdots&\vdots& &\vdots&\vdots&\vdots& &\vdots\\
\beta^{(\delta-1)t_{1,1}}&\beta^{(\delta-1)t_{1,2}}&\cdots&\beta^{(\delta-1)t_{1,\delta}}&0&0&\cdots&0\\
\hline
0&0&\cdots&0&\beta^{t_{2,1}}&\beta^{t_{2,2}}&\cdots&\beta^{t_{2,\delta}}\\
0&0&\cdots&0&\beta^{2t_{2,1}}&\beta^{2t_{2,2}}&\cdots&\beta^{2t_{2,\delta}}\\
\vdots&\vdots& &\vdots&\vdots&\vdots& &\vdots\\
0&0&\cdots&0&\beta^{(\delta-1)t_{2,1}}&\beta^{(\delta-1)t_{2,2}}&\cdots&\beta^{(\delta-1)t_{2,\delta}}\\
\hline
1&1&\cdots&1&1&1&\cdots&1\\
\gamma^{t_{1,1}m+i_1}&\gamma^{t_{1,2}m+i_1}&\cdots&\gamma^{t_{1,\delta}m+i_1}&\gamma^{t_{2,1}m+i_2}&\gamma^{t_{2,1}m+i_2}&\cdots&\gamma^{t_{2,\delta}m+i_2}\\
\end{array}},
\end{equation*}
has full rank.
\end{lemma}

\begin{IEEEproof}
Assume to the contrary that there exists $0\neq E=(e_{1,1},e_{1,2},\cdots,e_{1,\delta-1},e_{2,1},e_{2,2},\cdots,e_{2,\delta-1},e_0,e_{\delta})\in \F^{2\delta}_{q^b}$ satisfying $EM=0$.
Hence, the polynomials $f_1(x)=e_0+\gamma^{i_1}e_{\delta}x^{\delta}+\sum_{i=1}^{\delta-1}e_{1,i}x^i$ and $f_2(x)=e_0+e_{\delta}\gamma^{i_2}x^{\delta}+\sum_{i=1}^{\delta-1}e_{2,i}x^i$
have roots $\{\beta^{t_{1,i}}~:~1\leq i\leq \delta\}$ and $\{\beta^{t_{2,i}}~:~1\leq i\leq \delta\}$,
respectively. Note that $E\ne 0$ implies that $e_{\delta}\ne 0$, for otherwise, $\deg(f_1(x))<\delta$ and $\deg(f_2(x))<\delta$, but they each have $\delta$ distinct roots, a contradiction. By Vieta's formula,
we have
\[\frac{\prod_{1\leq i\leq \delta}\beta^{t_{1,i}}}{\prod_{1\leq i\leq \delta}\beta^{t_{2,i}}}=\gamma^{i_2-i_1}.\]
Hence, there exists an integer $t$ such that
$\beta^t=\gamma^{i_2-i_1}$, i.e., $\alpha^{mt}=\alpha^{\delta(i_2-i_1)}$.
It follows that
\[  mt \equiv \delta(i_2-i_1) \pmod{ma},\]
and then
\[ 0 \equiv \delta(i_2-i_1) \pmod{m}.\]
This contradicts the facts that $i_1\ne i_2$ and $\gcd(\delta,m)=1$, and
completes the proof.
\end{IEEEproof}

We can now prove that the constructed code is indeed a cyclic MR code.

\begin{theorem}
  \label{th:CPMDS}
  Assume the setting and notation of Construction~\ref{cons_CPMDS}.
  Then the code $\cC$ of Construction~\ref{cons_CPMDS} is a cyclic
  $(n=q^b-1,r,h=2,\delta,q^b)$-MR code, equivalently, a cyclic MR code
  with parameters $[n=q^b-1,k=mr-2,d]_{q^b}$ with repair sets of size
  $r+\delta-1$, and
  \[ d=\begin{cases}
  \delta+2 & r>2,\\
  2\delta+1 & r=2.
  \end{cases}\]
\end{theorem}

\begin{IEEEproof}
  Denote $\beta=\alpha^{m}$, and $\gamma=\alpha^{\delta}$. In the first
  step of our proof we contend that the following matrix is a
  parity-check matrix of $\cC$,
  \[H=
\parenv{\begin{array}{cccc|cccc|c|cccc}
1&0&\cdots &0&\beta&0&\cdots &0&\cdots&\beta^{a-1}&0&\cdots &0\\
1&0&\cdots &0&\beta^2&0&\cdots &0&\cdots&\beta^{2(a-1)}&0&\cdots &0\\
\vdots&\vdots& &\vdots&\vdots&\vdots& &\vdots& &\vdots&\vdots& &\vdots\\
1&0&\cdots &0&\beta^{\delta-1}&0&\cdots &0&\cdots&\beta^{(\delta-1)(a-1)}&0&\cdots &0\\
\hline
0&1&\cdots &0&0&\beta&\cdots &0&\cdots&0&\beta^{a-1}&\cdots &0\\
0&1&\cdots &0&0&\beta^2&\cdots &0&\cdots&0&\beta^{2(a-1)}&\cdots &0\\
\vdots&\vdots& &\vdots&\vdots&\vdots& &\vdots& &\vdots&\vdots& &\vdots\\
0&1&\cdots &0&0&\beta^{\delta-1}&\cdots &0&\cdots&0&\beta^{(\delta-1)(a-1)}&\cdots &0\\
\hline
\vdots & \vdots & & \vdots & \vdots & \vdots & & \vdots & & \vdots & \vdots & & \vdots \\
\hline
0&0&\cdots &1&0&0&\cdots &\beta&\cdots&0&0&\cdots &\beta^{a-1}\\
0&0&\cdots &1&0&0&\cdots &\beta^2&\cdots&0&0&\cdots &\beta^{2(a-1)}\\
\vdots&\vdots& &\vdots&\vdots&\vdots& &\vdots& &\vdots&\vdots& &\vdots\\
0&0&\cdots &1&0&0&\cdots &\beta^{\delta-1}&\cdots&0&0&\cdots &\beta^{(\delta-1)(a-1)}\\
\hline
1&1&\cdots &1&1&1&\cdots &1&\cdots&1&1&\cdots &1\\
1&\gamma&\cdots &\gamma^{m-1}&\gamma^{m}&\gamma^{m+1}&\cdots &\gamma^{2m-1}&\cdots&\gamma^{(a-1)m}&\gamma^{(a-1)m+1}&\cdots &\gamma^{n-1}\\
\end{array}}.
\]
  Define the following polynomial,
  \[f(x)\eqdef\prod_{i=1}^{m-1}(x-\alpha^{ai})=\sum_{j=0}^{m-1}e_jx^{j}.\]
  Clearly, $f(1)\neq 0$. We then have
  \begin{align*}
    &(e_0,e_1,\cdots,e_{m-1})
    \begin{pmatrix}
      1 &1 &1 &\cdots &1\\
      1 &\alpha^{a} &\alpha^{2a} &\cdots &\alpha^{a(n-1)}\\
      1 &\alpha^{2a} &\alpha^{4a} & \cdots &\alpha^{2a(n-1)}\\
      \vdots &\vdots &\vdots &  &\vdots\\
      1 &\alpha^{a(m-1)} &\alpha^{2a(m-1)} & \cdots &\alpha^{a(m-1)(n-1)}\\
    \end{pmatrix}\\
    &\quad =(f(1),0,\dots,0,f(\alpha^{am}),0,\dots,0,f(\alpha^{2am}),\dots,0,f(\alpha^{(a-1)am}),0,\dots,0)\\
    &\quad =(f(1),0,\dots,0,f(1),0,\dots,0,f(1),0,\dots,0,f(1),0,\dots,0).
  \end{align*}
  The preceding equation also means that for all $1\leq i\leq \delta-1$,
  \begin{align}
    &(e_0,e_1,\cdots,e_{m-1})
    \begin{pmatrix}
      1 &\alpha^{i} &\alpha^{2i} & \cdots &\alpha^{i(n-1)}\\
      1 &\alpha^{a+i} &\alpha^{2a+2i} & \cdots &\alpha^{a(n-1)+i(n-1)}\\
      1 &\alpha^{2a+i} &\alpha^{4a+2i} &\cdots &\alpha^{2a(n-1)+i(n-1)}\\
      \vdots &\vdots &\vdots & &\vdots\\
      1 &\alpha^{a(m-1)+i} &\alpha^{2a(m-1)+2i} & \cdots &\alpha^{a(m-1)(n-1)+i(n-1)}\\
    \end{pmatrix}\nonumber\\
    &\quad=(f(1),0,\dots,0,\alpha^{mi}f(1),0,\dots,0,\alpha^{2mi}f(1),0,\dots,0,\alpha^{(a-1)mi}f(1),0,\dots,0). \label{eq:dual}
  \end{align}

  Assume $G$ is a generator matrix for $\cC$. Recall that the roots of
  $\cC$ are $R=\mathset{\alpha^{ja+t}~:~1\leq j\leq m,1\leq t \leq
    \delta-1}\cup\mathset{1,\alpha^{\delta}}$. Hence, for all
  $1\leq j\leq m$ and $1\leq t\leq \delta-1$,
  \[ G\cdot\parenv{1, \alpha^{ja+t}, \alpha^{2(ja+t)},\dots,\alpha^{(n-1)(ja+t)}}^\T = 0.\]
  Define, for all $1\leq i\leq \delta-1$,
  \[ c_i = (1,0,\cdots,0,\beta^i,0,\cdots,0,\beta^{2i},\cdots,0,\beta^{(a-1)i},0,\cdots,0).\]
  Note that $c_i$ is a linear combination of the rows of matrix in \eqref{eq:dual}.
  Thus, the facts $\beta=\alpha^m$, and $f(1)\neq 0$ hint
  \[ G\cdot c_i^\T=0,\]
  and so, $c_i\in\cC^\perp$. Combining this with the fact that $\cC$
  is cyclic (and therefore, also $\cC^\perp$),
  $\sigma^j(c_i)\in\cC^\perp$ for all $j$, where we recall that
  $\sigma$ is the cyclic left-shift operator. Thus, the first
  $m(\delta-1)$ rows of $H$ contain codewords of $\cC^\perp$.  The
  remaining last two rows of $H$ correspond to parity checks for the
  roots $1$ and $\gamma=\alpha^\delta$, both of which are roots of
  $\cC$. If we now denote by $\cC'$ the $[n,k',d']_{q^b}$ code whose
  parity-check matrix is $H$, we can say $\cC\subseteq\cC'$. It
  remains to show that $\cC=\cC'$ to complete the proof.

  We first observe that since $H$ has $m(\delta-1)+2$ rows,
  \begin{equation}
    \label{eq:kprime}
    \dim(\cC')=k'\geq n-m(\delta-1)-2=mr-2.
  \end{equation}
  An inspection of $H$ reveals that $\cC'$ has all-symbol
  $(r,\delta)$-locality and the repair sets are given by $G_i\eqdef
  \angleenv{m}+i$ for $i\in[m]$. Plugging~\eqref{eq:kprime} into
  Theorem~\ref{lemma_bound_i} we obtain that the minimum distance of
  $\cC'$ satisfies
  \begin{equation}
    \label{eq:dprime}
    d' \leq \begin{cases}
      \delta+2 & r>2,\\
      2\delta+1 & r=2.
    \end{cases}
  \end{equation}

  Let us first handle the case of $r>2$. We contend that in that case,
  the minimum distance of $\cC'$ is at least $d'\geq \delta+2$. Even
  if we ignore the two bottom rows of $H$, the
  $(\delta-1)\times(r+\delta-1)$ Vandermonde matrices in the columns
  corresponding to a repair set show that any $\delta-1$ columns of
  $H$ are linearly independent. Thus, a linearly dependent set of
  columns from $H$ requires at least $\delta$ columns from each repair
  set it intersects. If we try to pick linearly dependent columns from
  a single repair set, then taking into account also the bottom two
  rows of $H$, the columns of a repair set also form a
  $(\delta+1)\times(r+\delta-1)$ Vandermonde matrix (recall that
  $\gamma^m=\beta^\delta$), and so $\delta+1$ columns of $H$ from
  the same repair set are still linearly independent. If instead we
  pick columns from more than one repair set, at least $2\delta$
  columns are required. Combined together, since $\delta\geq 2$, the
  smallest set of linearly dependent columns of $H$ contains at least
  $\delta+2$ columns, i.e., $d'\geq \delta+2$ as claimed. Together
  with~\eqref{eq:dprime},
  \[ d'=\delta+2.\]
  Again by Theorem~\ref{lemma_bound_i}, necessarily
  \[ k'= n-m(\delta-1)-2=mr-2.\]
  Finally, we note that
  \[ \dim(\cC)=k=n-\abs{R}=n-m(\delta-1)-2=k'=\dim(\cC').\]
  Since $\cC\subseteq\cC'$, and they are of equal dimension, we have
  $\cC=\cC'$, and $H$ is a parity-check matrix for $\cC$.

  We turn to the case of $r=2$. As in the previous case, a linearly
  dependent set of columns from $H$ requires at least $\delta$ columns
  from each repair set it intersects. However, this time, since $r=2$,
  we cannot choose $\delta+2$ columns from the same repair set, since
  each repair set contains exactly $\delta+1$ columns. Thus, a set of
  linearly dependent columns of $H$ contains at least $\delta$ columns
  each from two repair sets. However, by Lemma~\ref{lemma_full_rank},
  exactly $\delta$ columns each from two repair sets, still forms a
  set of linearly independent vectors. Thus, at least $2\delta+1$
  columns are required for a dependent set, namely, $d'\geq
  2\delta+1$. As in the previous case, by~\eqref{eq:dprime} we have
  \[ d'=2\delta+1,\]
  and then
  \[ k'=n-m(\delta-1)-2=k,\]
  and $\cC=\cC'$, as desired.

  In summary, we just proved the code $\cC$ is an optimal LRC which is
  cyclic. The fact that it is a $(q^b-1,r,2,\delta,q^b)$-MR code
  follows directly from Lemma~\ref{lemma_full_rank}, since any erasure
  pattern hitting two repair sets with $\delta$ erasures each,
  corresponds to a full-rank set of columns from $H$, and is therefore
  correctable.
\end{IEEEproof}

The cyclic MR codes by Construction~\ref{cons_CPMDS} have optimal
Hamming distance, and order-optimal field size with respect to the
bound in Lemma~\ref{lemma_lower_bound_F}-(1), where we consider
$\delta\geq 2$ as a constant. However, we can do better than that
when $r=2$, according to Corollary~\ref{Lemma_length_r=2}.

\begin{corollary}
When $r=2$, the cyclic MR codes generated by
Construction~\ref{cons_CPMDS} have optimal field size by
Corollary~\ref{Lemma_length_r=2}, provided that neither $q^b-1$, nor $q^b-2$,
are prime powers. When $r>2$, the cyclic MR codes by
Construction~\ref{cons_CPMDS} have optimal Hamming distance, and
order-optimal field size with respect to the bound in
Lemma~\ref{lemma_lower_bound_F}-(1), where we consider $\delta\geq 2$
as a constant.
\end{corollary}

\subsection{Turning Non-cyclic Codes into Cyclic Codes}

Previous works that constructed \emph{non-cyclic}
$(n,r,h,\delta,q)$-MR codes, for $h=2$, did so with
$q=\Theta(n(\delta-1))$ in~\cite{blaum2016construction}, and later,
with $q=\Theta(n)$~\cite{gopi2020maximally} (see
also~\cite{hu2016new}, that obtained $q=\Theta(n)$ for the special
case of $n=2(r+\delta-1)$). Of particular interest to us are the
$(n,r,2,\delta,q^b)$-MR codes from~\cite[Theorem
  IV.2]{gopi2020maximally}. These MR codes have the same parameters as
Construction~\ref{cons_CPMDS}. However, they are not cyclic MR codes
directly. In what follows, we shall attempt to determine whether the
MR codes generated in~\cite[Theorem IV.2]{gopi2020maximally} can be
rearranged to become cyclic codes. Along the way, we shall prove some
interesting facts concerning cyclic optimal LRCs.

As a first step, we show the repair sets of cyclic optimal LRCs are
severely restricted.

\begin{theorem}\label{theorem_repair_set_cyc}
  Let $\cC$ be a cyclic optimal LRC with parameters $[n,k,d]_q$ and
  all-symbol $(r,\delta)$-locality. Write $k=ur+v$ with $0<v\leq r$.
  If $u\geq 2(r-v+1)$, then for any repair set $S\subseteq \Z_n$, and
  any $j\in\Z_n$, either $S+j=S$ or $(S+j)\cap S=\emptyset$.
\end{theorem}

The technical proof of Theorem~\ref{theorem_repair_set_cyc} is
deferred to the appendix. As an immediate consequence, we now show
that the repair sets of cyclic optimal LRCs must be cosets of $\Z_n$.

\begin{corollary}\label{corollary_repair_set}
  Let $\cC$ be a cyclic optimal LRC with parameters $[n,k,d]_q$ and
  all-symbol $(r,\delta)$-locality (where, to avoid trivial cases, we
  assume that $\cC$ does not have all-symbol $(r-1,\delta)$-locality).
  Let $k=ur+v$, $0<v\leq r$. If $u\geq 2(r-v+1)$, then
  $n=m(r+\delta-1)$, $m\in\N$, and the repair sets of $\cC$ are
  \[ G_i\eqdef \angleenv{m}+i = \mathset{ jm + i ~:~ j\in\Z } \subseteq\Z_n,\]
  for all $i\in\Z$.
\end{corollary}

\begin{IEEEproof}
Let $S_0\subseteq\Z_n$ be a repair set such that $0\in S_0$. By
Theorem~\ref{theorem_repair_set_cyc} we have $S_0+i=S_0\subseteq \Z_n$
for any $i\in S_0$. Thus, $S_0$ is a subgroup of the cyclic group
$(\Z_{n},+)$. Note that $|S_0|\leq r+\delta-1$.  If
$|S_0|<r+\delta-1$, then the fact $S_0+i$, for $i\in \Z_n$, are also
repair sets for $\cC$, implies that $\cC$ has all-symbol
$(r-1,\delta)$-locality, which contradicts our assumption. Thus,
$|S_0|=r+\delta-1$, $S_0=\angleenv{m}$, and $(r+\delta-1)|n$.

Let $S$ be any repair set of $\cC$. the same analysis shows that
$|S|=r+\delta-1$. Note that $S-i$ for any $i\in S$ is still a repair
set of $\cC$. Now it is easy to check that $S-i$ is a
$r+\delta-1$-subgroup of $(\Z_n,+)$, i.e., $S-i=S_0$, which completes
the proof.
\end{IEEEproof}

\begin{remark}
  Corollary~\ref{corollary_repair_set} shows that the condition
  $(r+\delta-1)|n$ is not a restriction when $u\geq 2(r-v+1)$, but
  rather a consequence.
\end{remark}

\begin{remark}
  For the case $u=1$ (i.e., $k=r+v$), and $(r+\delta-1)\nmid n$,
  explicit constructions were proposed in~\cite[Corollaries
    27, 37, 43]{qiu2020new} for cyclic optimal
  LRCs. Corollary~\ref{corollary_repair_set} implies that
  constructions with such parameters are possible only if
  $1=u<2(r-v+1)$, i.e., $r\geq v$.
\end{remark}

Further building on Corollary~\ref{corollary_repair_set}, we can now
show that cyclic optimal LRCs have a parity-check matrix with a nice
form.

\begin{corollary}\label{corollary_ parity}
  Let $\cC$ be a cyclic optimal LRC with parameters $[n,k,d]_q$, and
  all-symbol $(r,\delta)$-locality (where, to avoid trivial cases, we
  assume that $\cC$ does not have all-symbol $(r-1,\delta)$-locality).
  If $u\geq 2(r-v+1)$, then a parity-check matrix of $\cC$ can be
  given in the following form:
\[
H=\parenv{\begin{array}{cccc|cccc|c|cccc}
s_0&0&\cdots &0&s_1&0&\cdots &0&\cdots&s_{a-1}&0&\cdots &0\\
0&s_0&\cdots &0&0&s_1&\cdots &0&\cdots&0&s_{a-1}&\cdots &0\\
\vdots&\vdots& \ddots  &\vdots&\vdots&\vdots& \ddots &\vdots& &\vdots&\vdots& \ddots &\vdots\\
0&0&\cdots &s_0&0&0&\cdots &s_1&\cdots&0&0&\cdots &s_{a-1}\\
h_0&h_1&\cdots &h_{m-1}&h_{m}&h_{m+1}&\cdots &h_{2m-1}&\cdots&h_{(a-1)m}&h_{(a-1)m+1}&\cdots &h_{am-1}\\
\end{array}},
\]
where $s_i,h_j$ are column vectors, $(s_0,s_2,\cdots,s_{a-1})$ is a
parity-check matrix of a cyclic code with minimum Hamming distance of
at least $\delta$, $a=r+\delta-1$, $m=n/a$, and $(h_0,h_1,\cdots,
h_{am-1})$ corresponds to the global parity checks. Moreover, the
punctured codes satisfy $\cC|_{G_i}=\cC|_{G_0}$ for all $i\in[m]$,
where $G_i\eqdef\angleenv{m}+i$. Similarly, the shortened codes
satisfy $\cC|^{G_i}=\cC|^{G_0}$ for all $i\in[m]$, where $\cC|^{G_i}$
is the code whose parity check matrix contains only the columns
corresponding to $G_i$ from $H$.
\end{corollary}
\begin{IEEEproof}
The parity-check matrix follows directly from
Corollary~\ref{corollary_repair_set} and the fact that
$\cC|_{\angleenv{m}}=\cC|_{\angleenv{m}+i}$ for $i\in \Z$ is also a
cyclic code. Additionally, since $\cC$ is cyclic, trivially we have
$\cC|_{G_i}=\cC|_{G_0}$ and $\cC|^{G_i}=\cC|^{G_0}$.
\end{IEEEproof}

Now we recall a construction, which was first introduced
in~\cite{gopi2020maximally}.

\begin{construction}[\cite{gopi2020maximally}]\label{cons_MR_known}
Let $q$ be a prime power, $b\in\N$, $n=q^b-1$, $a=r+\delta-1$,
$a|(q-1)$, and $m=n/a$.  Let $\alpha$ be a primitive element of
$\F_{q^b}$, $\beta=\alpha^{m}$, and $\lambda=\alpha^s$, $\gcd(s,m)=1$. The following
parity-check matrix defines an $(n,r,2,\delta,q^b)$-MR code,
\begin{equation*}
H=\parenv{\begin{array}{cccc|cccc|c|cccc}
1&0&\cdots &0&\beta&0&\cdots &0&\cdots&\beta^{a-1}&0&\cdots &0\\
1&0&\cdots &0&\beta^2&0&\cdots &0&\cdots&\beta^{2(a-1)}&0&\cdots &0\\
\vdots&\vdots& &\vdots&\vdots&\vdots& &\vdots& &\vdots&\vdots& &\vdots\\
1&0&\cdots &0&\beta^{\delta-1}&0&\cdots &0&\cdots&\beta^{(\delta-1)(a-1)}&0&\cdots &0\\
0&1&\cdots &0&0&\beta&\cdots &0&\cdots&0&\beta^{a-1}&\cdots &0\\
\vdots&\vdots& &\vdots&\vdots&\vdots& &\vdots& &\vdots&\vdots& &\vdots\\
0&1&\cdots &0&0&\beta^{\delta-1}&\cdots &0&\cdots&0&\beta^{(\delta-1)(a-1)}&\cdots &0\\
\vdots&\vdots&\ddots &\vdots&\vdots&\vdots&\ddots &\vdots& &\vdots&\vdots&\ddots &\vdots\\
0&0&\cdots &1&0&0&\cdots &\beta&\cdots&0&0&\cdots &\beta^{a-1}\\
\vdots&\vdots& &\vdots&\vdots&\vdots& &\vdots& &\vdots&\vdots& &\vdots\\
0&\cdots&\cdots &1&0&0&\cdots &\beta^{\delta-1}&\cdots&0&0&\cdots &\beta^{(\delta-1)(a-1)}\\
\lambda^0&\lambda^1&\cdots &\lambda^{m-1}&\lambda^0&\lambda^1&\cdots &\lambda^{m-1}&\cdots&\lambda^0&\lambda^1&\cdots &\lambda^{m-1}\\
1&1&\cdots &1&\beta^{\delta}&\beta^{\delta}&\cdots &\beta^{\delta}&\cdots&\beta^{\delta(a-1)}&\beta^{\delta(a-1)}&\cdots &\beta^{\delta(a-1)}\\
\end{array}}.
\end{equation*}
\end{construction}

To simply our notation, we define,
\[\vv{x} \eqdef \begin{pmatrix}
  x \\ x^2 \\ \vdots \\ x^{\delta-1}
\end{pmatrix}.\]
In this notation, the matrix $H$ from Construction~\ref{cons_MR_known}
becomes,
\[ H = \parenv{ \begin{array}{cccc|cccc|c|cccc}
    \vv{1} & \vv{0} & \cdots & \vv{0} & \vv{\beta} & \vv{0} & \cdots & \vv{0} & \cdots & \vv{\beta^{a-1}} & \vv{0} & \cdots & \vv{0} \\
    \vv{0} & \vv{1} & \cdots & \vv{0} & \vv{0} & \vv{\beta} & \cdots & \vv{0} & \cdots & \vv{0} & \vv{\beta^{a-1}} & \cdots & \vv{0} \\
    \vdots & \vdots & \ddots & \vdots & \vdots & \vdots & \ddots & \vdots & & \vdots & \vdots & \ddots & \vdots \\
    \vv{0} & \vv{0} & \cdots & \vv{1} & \vv{0} & \vv{0} & \cdots & \vv{\beta} & \cdots & \vv{0} & \vv{0} & \cdots & \vv{\beta^{a-1}} \\
\lambda^0&\lambda^1&\cdots &\lambda^{m-1}&\lambda^0&\lambda^1&\cdots &\lambda^{m-1}&\cdots&\lambda^0&\lambda^1&\cdots &\lambda^{m-1}\\
1&1&\cdots &1&\beta^{\delta}&\beta^{\delta}&\cdots &\beta^{\delta}&\cdots&\beta^{\delta(a-1)}&\beta^{\delta(a-1)}&\cdots &\beta^{\delta(a-1)}\\
\end{array}} \]

One cannot avoid seeing a similarity between the parity-check matrix
of Construction~\ref{cons_MR_known}, and the parity-check matrix found
in Theorem~\ref{th:CPMDS} for the code from
Construction~\ref{cons_CPMDS}. However, the code from
Construction~\ref{cons_MR_known} is not cyclic, but rather
quasi-cyclic. In what follows we study whether permuting it produces a
cyclic code.

Let $\bS_n$ denote the set of permutations over $\Z_n$, for any
$n\in\N$. Each permutation in $\bS_n$ may be thought of as a bijection
in $\Z_n^{\Z_n}$, namely, a bijection from $\Z_n$ to $\Z_n$. Let $\cC$
be a code of length $n$, whose coordinates are indexed by $\Z_n$. If
$\ell\in\bS_n$ is a permutation, we define the permutation of $\cC$ by
$\ell$ as
\[\cC_{\ell}\eqdef\mathset{(c_{\ell(0)},c_{\ell(1)},\dots,c_{\ell(n-1)}) ~:~ (c_0,c_1,\dots,c_{n-1})\in \cC)}. \]
If $\cC$ is a cyclic code, it is natural to ask for what permutations
$\ell\in\bS_n$, $\cC_\ell$ is also cyclic. Apart from the trivial
cyclic shifts of $\cC$, a natural subset of candidate permutations are
\emph{multipliers}, namely,
\begin{align*}
  \mu_{t}(x) &\eqdef xt \bmod n, \\
  \euler{n} &\eqdef \mathset{ 1\leq t\leq n ~:~ \gcd(t,n)=1},\\
  \Upsilon(n) &\eqdef \mathset{ \mu_{t} ~:~ t\in \euler{n}}.
\end{align*}
P\'alfy~\cite{palfy1987isomorphism} proved that, in many cases, multipliers
are the essential permutations keeping a code cyclic:

\begin{lemma}[\cite{palfy1987isomorphism}]\label{lemma_multipliers}
  Consider codes of length $n$ whose coordinates are indexed by
  $\Z_n$.
  \begin{enumerate}
  \item
    When $\gcd(n,\varphi(n))=1$ or $n=4$, for all cyclic codes $\cC$,
    if $\cC_{\ell'}$, $\ell'\in\bS_n$, is also a cyclic code, then
    there is a multiplier $\ell\in\Upsilon(n)$ such that
    $\cC_{\ell'}=\cC_\ell$.
  \item
    When $\gcd(n,\varphi(n))\neq 1$ and $n\neq 4$, there exists a
    cyclic code $\cC$, and $\ell'\in\bS_n$ such that $\cC_{\ell'}$ is
    cyclic, but $\cC_{\ell'}\neq \cC_{\ell}$ for all multipliers
    $\ell\in\Upsilon(n)$.
  \end{enumerate}
  Here $\varphi(\cdot)$ denotes Euler's totient function.
\end{lemma}

Drawing inspiration from Lemma~\ref{lemma_multipliers}, we address the
(different) question of finding permutations from $\bS_n$ that turn
the non-cyclic code of Construction~\ref{cons_MR_known} into a cyclic
code. Recall that in the setting of Construction~\ref{cons_MR_known},
$a,m,n\in\N$, and $n=ma$. We now define a set of functions from $\Z_n$
to $\Z_n$ as follows:
\[ \mu_{t,z}(xm+i) \eqdef \parenv{xmt_i + z_i} \bmod n,\]
where we assume $x\in[a]$, $i\in[m]$, $t=(t_0,\dots,t_{m-1})\in\Z^m$,
and $z=(z_0,\dots,z_{m-1})\in\Z^m$. We then define the set,
\[ \Psi(n,a) \eqdef \mathset{ \mu_{t,z} ~:~ t\in(\euler{a})^m, z\in\Z^m, (z\bmod m)\in\bS_m},\]
and where by abuse of notation, $z\bmod m$ denotes the $\Z_m\to\Z_m$
mapping that maps $i\mapsto (z_i \bmod m)$.

We would like to make some easy observations concerning the elements
of $\Psi(n,a)$. Denote $G_0\eqdef\angleenv{m}\subseteq \Z_n$. Then
$G_0$ is an additive subgroup of $\Z_n$, and $G_0\cong \Z_a$. Let us
denote the cosets of $G_0$ by $G_i\eqdef G_0+i$, for all $i\in\Z$. We
now note that $j\mapsto jt \bmod n$ is a bijection from $G_0$ to $G_0$
if and only if $\gcd(t,a)=1$. Thus, $\ell_{t,z}|_{G_i}$ (i.e., the
restriction of $\ell_{t,z}$ to $G_i$) is a bijection from $G_i$ to
$G_{z_i \bmod m}$. With the extra requirement that $(z \bmod
m)\in\bS_m$, we have that distinct cosets $G_i$ are mapped to distinct
cosets $G_{z_i \bmod m}$, and hence, $\Psi(n,a)\subseteq\bS_n$,
namely, $\Psi$ comprises of permutations over $\Z_n$.

\begin{theorem}\label{th:multi_permu}
  Assume the notation and setting of Construction~\ref{cons_MR_known},
  and let $\cC$ be the resulting code when $r\geq 3$. Then there exists a
  permutation $\ell\in \Psi(n,a)$ such that $\cC_\ell$ is a cyclic
  code if and only if $\gcd(m,\frac{a}{\gcd(a,\delta)})=1$.
\end{theorem}

\begin{IEEEproof}
  We first observe that $\gcd(m,\frac{a}{\gcd(a,\delta)})=1$ if and
  only if the equation $\delta m \tau \equiv \delta \pmod{a}$ has at
  least one solution $\tau\in\Z_a$. We now prove both directions of
  the claim.

  In the first direction, assume $\delta m\tau\equiv \delta\pmod{a}$
  has a solution $\tau\in\Z_a$. Consider the permutation
  $\ell=\ell_{t,z}\in\Psi(n,a)$ for which $t=(1,\dots,1)$, and
  $z=(z_0,\dots,z_{m-1})$, where $z_i = i+m\tau i$. Applying $\ell$ to
  the coordinates of $\cC$, the parity-check matrix $H$ from
  Construction~\ref{cons_MR_known} becomes
\[
H_\ell=
\parenv{\begin{array}{cccc|cccc|c|ccc}
\vv{1}&\vv{0}&\cdots &\vv{0}&\vv{\beta}&\vv{0}&\cdots &\vv{0}&\cdots&\vv{\beta^{a-1}}&\cdots &\vv{0}\\
\vv{0}&\vv{\beta^\tau}&\cdots &\vv{0}&\vv{0}&\vv{\beta^{\tau+1}}&\cdots &\vv{0}&\cdots&\vv{0}&\cdots &\vv{0}\\
\vdots&\vdots&\ddots &\vdots&\vdots&\vdots&\ddots &\vdots& &\vdots&\ddots &\vdots\\
\vv{0}&\vv{0}&\cdots &\vv{\beta^{\tau(m-1)}}&\vv{0}&\vv{0}&\cdots &\vv{\beta^{\tau(m-1)+1}}&\cdots&\vv{0}&\cdots &\vv{\beta^{\tau(m-1)+a-1}}\\
\lambda^0&\lambda^1&\cdots &\lambda^{m-1}&\lambda^0&\lambda^1&\cdots &\lambda^{m-1}&\cdots&\lambda^0&\cdots &\lambda^{m-1}\\
1&\beta^{\tau\delta}&\cdots &\beta^{\tau(m-1)\delta}&\beta^{\tau m\delta}&\beta^{\tau(m+1)\delta}&\cdots &\beta^{\tau(2m-1)\delta}&\cdots&\beta^{\tau m(a-1)\delta}&\cdots &\beta^{\tau(am-1)\delta}\\
\end{array}}.
\]
which is a parity-check matrix for $\cC_\ell$. Here we used $\delta
m\tau\equiv \delta\pmod{a}$ to get that
$\beta^{\delta m \tau}=\beta^\delta$. Now, by dividing some of the rows with
appropriate scalars, another parity-check matrix for $\cC_\ell$ is the
following:
\[
H'_\ell=
\parenv{\begin{array}{cccc|cccc|c|ccc}
\vv{1}&\vv{0}&\cdots &\vv{0}&\vv{\beta}&\vv{0}&\cdots &\vv{0}&\cdots&\vv{\beta^{a-1}}&\cdots &\vv{0}\\
\vv{0}&\vv{1}&\cdots &\vv{0}&\vv{0}&\vv{\beta}&\cdots &\vv{0}&\cdots&\vv{0}&\cdots &\vv{0}\\
\vdots&\vdots&\ddots &\vdots&\vdots&\vdots&\ddots &\vdots& &\vdots&\ddots &\vdots\\
\vv{0}&\vv{0}&\cdots &\vv{1}&\vv{0}&\vv{0}&\cdots &\vv{\beta}&\cdots&\vv{0}&\cdots &\vv{\beta^{a-1}}\\
\lambda^0&\lambda^1&\cdots &\lambda^{m-1}&\lambda^0&\lambda^1&\cdots &\lambda^{m-1}&\cdots&\lambda^0&\cdots &\lambda^{m-1}\\
1&\beta^{\tau\delta}&\cdots &\beta^{\tau(m-1)\delta}&\beta^{\tau m\delta}&\beta^{\tau(m+1)\delta}&\cdots &\beta^{\tau(2m-1)\delta}&\cdots&\beta^{\tau m(a-1)\delta}&\cdots &\beta^{\tau(am-1)\delta}\\
\end{array}}.
\]
It is now clear that $\cC_\ell$ is cyclic, since $H'_\ell c^\T=0$
implies $H'_\ell (\sigma(c))^\T = 0$, i.e., $c\in \cC_\ell$ implies
$\sigma(c)\in\cC_\ell$.

In the second direction, assume that there exists $\ell\in \Psi(n,a)$
such that $\cC_{\ell}$ is cyclic. Assume to the contrary that $\delta
m \tau \not\equiv \delta\pmod{a}$ for all $\tau\in\Z_a$. Let us write
$\ell=\ell_{t,z}\in\Psi(n,a)$, with
$t=(t_0,\dots,t_{m-1})\in(\euler{a})^m$, and
$z=(z_0,\dots,z_{m-1})\in\Z^m$. We can now write,
\[ \ell(xm+i) = (xmt_i + z_i) \bmod n = ((x+\tau_i)mt_i + \zeta_i) \bmod n,\]
with $x,\tau_i\in[a]$, and $i,\zeta_i\in[m]$. Let us further define
$\beta_i\eqdef \beta^{t_i}$. We can now apply $\ell$ to the order of
the columns of $H$ from Construction~\ref{cons_MR_known} to obtain a
parity-check matrix $H_\ell$ for the code $\cC_\ell$. By rearranging
the order of the rows of the matrix, we may write,
\[
  H_\ell=
\parenv{\begin{array}{cccc|cccc|c|ccc}
\vv{\beta^{\tau_0}_0}&\vv{0}&\cdots &\vv{0}&\vv{\beta^{\tau_0+1}_0}&\vv{0}&\cdots &\vv{0}&\cdots&\vv{\beta_0^{\tau_0+a-1}}&\cdots &\vv{0}\\
\vv{0}&\vv{\beta^{\tau_1}_1}&\cdots &\vv{0}&\vv{0}&\vv{\beta_1^{\tau_1+1}}&\cdots &\vv{0}&\cdots&\vv{0}&\cdots &\vv{0}\\
\vdots&\vdots&\ddots &\vdots&\vdots&\vdots&\ddots &\vdots& &\vdots&\ddots &\vdots\\
\vv{0}&\vv{0}&\cdots &\vv{\beta_{m-1}^{\tau_{m-1}}}&\vv{0}&\vv{0}&\cdots &\vv{\beta_{m-1}^{\tau_{m-1}+1}}&\cdots&\vv{0}&\cdots &\vv{\beta_{m-1}^{\tau_{m-1}+a-1}}\\
\lambda^{\zeta_0}&\lambda^{\zeta_1}&\cdots &\lambda^{\zeta_{m-1}}&\lambda^{\zeta_0}&\lambda^{\zeta_1}&\cdots &\lambda^{\zeta_{m-1}}&\cdots&\lambda^{\zeta_0}&\cdots &\lambda^{\zeta_{m-1}}\\
\beta^{\tau_0\delta}_0&\beta_1^{\tau_1\delta}&\cdots &\beta_{m-1}^{\tau_{m-1}\delta}&\beta^{(\tau_0+1)\delta}_0
&\beta_1^{(\tau_1+1)\delta}&\cdots &\beta_{m-1}^{(\tau_{m-1}+1)\delta}&\cdots&\beta_0^{\delta (\tau_0+a-1)}&\cdots &\beta_{m-1}^{\delta(\tau_{m-1}+a-1)}\\
\end{array}}.
\]

Recall that, by construction, the multiplicative order of $\beta$ is
$o(\beta)=a$. Since $\gcd(t_j,a)=1$, we also have that $o(\beta_j)=a$,
for all $j\in\Z_m$. Taking into account that $r\geq 3$, namely,
$a=r+\delta-1\geq \delta+2$, we have that $1,\beta_j,
\beta_j^2,\dots,\beta_j^{\delta}$ are all distinct. Let us look at
the columns of $H_\ell$ that correspond to $G_j$ for some
$j\in\Z_m$. These columns, after removing all-zero rows, form a
(transposed) $(\delta+1)\times a$ Vandermonde matrix.
\begin{equation}
  \label{eq:gjv}
  \begin{pmatrix}
    \vv{\beta_j^{\tau_j}} & \vv{\beta_j^{\tau_j+1}} & \dots & \vv{\beta_j^{\tau_j+a-1}} \\
    \lambda^{\zeta_j} & \lambda^{\zeta_j} & \dots & \lambda^{\zeta_j} \\
    \beta_j^{\tau_j \delta} & \beta_j^{(\tau_j+1)\delta} & \dots & \beta_j^{(\tau_j+a-1)\delta}
  \end{pmatrix}=\Pi\cdot\diag(\beta_j^{\tau_j},\dots,\beta_j^{\tau_j(\delta-1)},\lambda^{\zeta_j},\beta_j^{\tau_j \delta})\cdot\begin{pmatrix}
  1 & 1 & \dots &1 \\ 1 & \beta_j & \dots & \beta_j^{a-1} \\ 1 &
  \beta_j^2 & \dots & \beta_j^{2(a-1)} \\ \vdots & \vdots & & \vdots
  \\ 1 & \beta_j^{\delta} & \dots &
  \beta_j^{\delta(a-1)}\end{pmatrix},
\end{equation}
where $\Pi$ is a permutation matrix that moves the second row from the
bottom to the top. Since $1,\beta_j,\beta_j^2,\dots,\beta_j^{\delta}$
are all distinct, the rows of~\eqref{eq:gjv} are linearly
independent. Thus, a linear combination of the rows of $H_\ell$ that
results in zeros in all the positions of $G_j$ must be a trivial
combination.

We now use the fact that $H_\ell$ is a parity-check matrix for a
cyclic code. By adding cyclic rotations of existing rows in $H_\ell$,
we obtain $H'_\ell$ which is also a parity-check matrix for the same
code,
\[
  H'_\ell=
\parenv{\begin{array}{cccc|cccc|c|ccc}
\vv{\beta^{\tau_i}_i}&\vv{0}&\cdots &\vv{0}&\vv{\beta^{\tau_i+1}_i}&\vv{0}&\cdots &\vv{0}&\cdots&\vv{\beta_i^{\tau_i+a-1}}&\cdots &\vv{0}\\ \hline
\vv{\beta^{\tau_0}_0}&\vv{0}&\cdots &\vv{0}&\vv{\beta^{\tau_0+1}_0}&\vv{0}&\cdots &\vv{0}&\cdots&\vv{\beta_0^{\tau_0+a-1}}&\cdots &\vv{0}\\
\vv{0}&\vv{\beta^{\tau_1}_1}&\cdots &\vv{0}&\vv{0}&\vv{\beta_1^{\tau_1+1}}&\cdots &\vv{0}&\cdots&\vv{0}&\cdots &\vv{0}\\
\vdots&\vdots&\ddots &\vdots&\vdots&\vdots&\ddots &\vdots& &\vdots&\ddots &\vdots\\
\vv{0}&\vv{0}&\cdots &\vv{\beta_{m-1}^{\tau_{m-1}}}&\vv{0}&\vv{0}&\cdots &\vv{\beta_{m-1}^{\tau_{m-1}+1}}&\cdots&\vv{0}&\cdots &\vv{\beta_{m-1}^{\tau_{m-1}+a-1}}\\
\lambda^{\zeta_0}&\lambda^{\zeta_1}&\cdots &\lambda^{\zeta_{m-1}}&\lambda^{\zeta_0}&\lambda^{\zeta_1}&\cdots &\lambda^{\zeta_{m-1}}&\cdots&\lambda^{\zeta_0}&\cdots &\lambda^{\zeta_{m-1}}\\
\beta^{\tau_0\delta}_0&\beta_1^{\tau_1\delta}&\cdots &\beta_{m-1}^{\tau_{m-1}\delta}&\beta^{(\tau_0+1)\delta}_0
&\beta_1^{(\tau_1+1)\delta}&\cdots &\beta_{m-1}^{(\tau_{m-1}+1)\delta}&\cdots&\beta_0^{\delta (\tau_0+a-1)}&\cdots &\beta_{m-1}^{\delta(\tau_{m-1}+a-1)}\\
\end{array}},
\]
where $i\in\Z_m$. However, these added rows must be linear
combinations of the rows of $H_\ell$. Since they contain zeros in all
the entries of $G_j$, $j\neq 0$, these linear combinations cannot use
the last two rows of $H_\ell$. It now follows that
\[ \rank
\begin{pmatrix}
  \vv{\beta_i^{\tau_i}} & \vv{\beta_i^{\tau_i+1}} & \dots & \vv{\beta_i^{\tau_i+a-1}} \\
  \vv{\beta_0^{\tau_0}} & \vv{\beta_0^{\tau_0+1}} & \dots & \vv{\beta_0^{\tau_0+a-1}} \\
\end{pmatrix}
= \rank
\begin{pmatrix}
  \vv{\beta_0^{\tau_0}} & \vv{\beta_0^{\tau_0+1}} & \dots & \vv{\beta_0^{\tau_0+a-1}}
\end{pmatrix}.
\]
After the same treatment as~\eqref{eq:gjv}, this gives
\[
\rank
\begin{pmatrix}
  1 & \beta_i & \dots & \beta_i^{a-1} \\
  1 & \beta_i^2 & \dots & \beta_i^{2(a-1)} \\
  \vdots & \vdots & & \vdots \\
  1 & \beta_i^{\delta-1} & \dots & \beta_i^{(\delta-1)(a-1)}\\
  1 & \beta_0 & \dots & \beta_0^{a-1} \\
  1 & \beta_0^2 & \dots & \beta_0^{2(a-1)} \\
  \vdots & \vdots & & \vdots \\
  1 & \beta_0^{\delta-1} & \dots & \beta_0^{(\delta-1)(a-1)}\\
\end{pmatrix}
=
\rank
\begin{pmatrix}
  1 & \beta_0 & \dots & \beta_0^{a-1} \\
  1 & \beta_0^2 & \dots & \beta_0^{2(a-1)} \\
  \vdots & \vdots & & \vdots \\
  1 & \beta_0^{\delta-1} & \dots & \beta_0^{(\delta-1)(a-1)}
\end{pmatrix}.
\]
If $\mathset{\beta_0,\beta_0^2,\dots,\beta_0^{\delta-1}}\neq
\mathset{\beta_i,\beta_i^2,\dots,\beta_i^{\delta-1}}$, then by the
fact that $r\geq 3$, we would have a contradiction to the rank
equality above. It follows that
\begin{equation}
  \label{eq:seqbeta}
  \mathset{\beta_0,\beta_0^2,\dots,\beta_0^{\delta-1}}=\mathset{\beta_i,\beta_i^2,\dots,\beta_i^{\delta-1}},
\end{equation}
for all $i\in\Z_m$.

We now repeat the argument, with an extra step. Take $H'_\ell$ and add to it
a cyclic rotation of its last row to obtain the following parity-check
matrix for the same code,
\[
  H''_\ell=
\parenv{\begin{array}{cccc|ccc|c|ccc}
\vv{\beta^{\tau_i}_i}&\vv{0}&\cdots &\vv{0}&\vv{\beta^{\tau_i+1}_i}&\cdots &\vv{0}&\cdots&\vv{\beta_i^{\tau_i+a-1}}&\cdots &\vv{0}\\ \hline
\vv{\beta^{\tau_0}_0}&\vv{0}&\cdots &\vv{0}&\vv{\beta^{\tau_0+1}_0}&\cdots &\vv{0}&\cdots&\vv{\beta_0^{\tau_0+a-1}}&\cdots &\vv{0}\\
\vv{0}&\vv{\beta^{\tau_1}_1}&\cdots &\vv{0}&\vv{0}&\cdots &\vv{0}&\cdots&\vv{0}&\cdots &\vv{0}\\
\vdots&\vdots&\ddots &\vdots&\vdots&\ddots &\vdots& &\vdots&\ddots &\vdots\\
\vv{0}&\vv{0}&\cdots &\vv{\beta_{m-1}^{\tau_{m-1}}}&\vv{0}&\cdots &\vv{\beta_{m-1}^{\tau_{m-1}+1}}&\cdots&\vv{0}&\cdots &\vv{\beta_{m-1}^{\tau_{m-1}+a-1}}\\
\lambda^{\zeta_0}&\lambda^{\zeta_1}&\cdots &\lambda^{\zeta_{m-1}}&\lambda^{\zeta_0}&\cdots &\lambda^{\zeta_{m-1}}&\cdots&\lambda^{\zeta_0}&\cdots &\lambda^{\zeta_{m-1}}\\
\beta^{\tau_0\delta}_0&\beta_1^{\tau_1\delta}&\cdots &\beta_{m-1}^{\tau_{m-1}\delta}&\beta^{(\tau_0+1)\delta}_0 &
\cdots &\beta_{m-1}^{(\tau_{m-1}+1)\delta}&\cdots&\beta_0^{\delta (\tau_0+a-1)}&\cdots &\beta_{m-1}^{\delta(\tau_{m-1}+a-1)}\\ \hline
\beta_i^{\tau_i\delta}&\beta_{i+1}^{\tau_{i+1}\delta}&\cdots &\beta_{i-1}^{(\tau_{i-1}+1)\delta}&\beta_i^{(\tau_i+1)\delta}
&\cdots &\beta_{i-1}^{(\tau_{i-1}+2)\delta}&\cdots&\beta_i^{\delta (\tau_i+a-1)}&\cdots &\beta_{i-1}^{\tau_{i-1}\delta}\\
\end{array}},
\]
where $i\in\Z_m$. Again, this added row is linearly dependent on the
others, and so, looking at the columns of $G_0$ we obtain the rank
equality
\[ \rank
\begin{pmatrix}
  \vv{\beta_i^{\tau_i}} & \vv{\beta_i^{\tau_i+1}} & \dots & \vv{\beta_i^{\tau_i+a-1}} \\
  \vv{\beta_0^{\tau_0}} & \vv{\beta_0^{\tau_0+1}} & \dots & \vv{\beta_0^{\tau_0+a-1}} \\
  \lambda^{\zeta_0} & \lambda^{\zeta_0} & \dots & \lambda^{\zeta_0} \\
  \beta_0^{\tau_0 \delta} & \beta_0^{(\tau_0+1)\delta} & \dots & \beta_0^{(\tau_0+a-1)\delta} \\
  \beta_i^{\tau_i \delta} & \beta_i^{(\tau_i+1)\delta} & \dots & \beta_i^{(\tau_i+a-1)\delta} \\
\end{pmatrix}
= \rank
\begin{pmatrix}
  \vv{\beta_0^{\tau_0}} & \vv{\beta_0^{\tau_0+1}} & \dots & \vv{\beta_0^{\tau_0+a-1}}\\
  \lambda^{\zeta_0} & \lambda^{\zeta_0} & \dots & \lambda^{\zeta_0} \\
  \beta_0^{\tau_0 \delta} & \beta_0^{(\tau_0+1)\delta} & \dots & \beta_0^{(\tau_0+a-1)\delta} \\
\end{pmatrix}.
\]
Again, using the same steps as in~\eqref{eq:gjv}, we get
\[
\rank
\begin{pmatrix}
  1 & 1 & \dots & 1 \\
  1 & \beta_i & \dots & \beta_i^{a-1} \\
  1 & \beta_i^2 & \dots & \beta_i^{2(a-1)} \\
  \vdots & \vdots & & \vdots \\
  1 & \beta_i^{\delta} & \dots & \beta_i^{\delta(a-1)}\\
  1 & \beta_0 & \dots & \beta_0^{a-1} \\
  1 & \beta_0^2 & \dots & \beta_0^{2(a-1)} \\
  \vdots & \vdots & & \vdots \\
  1 & \beta_0^{\delta} & \dots & \beta_0^{\delta(a-1)}\\
\end{pmatrix}
=
\rank
\begin{pmatrix}
  1 & 1 & \dots & 1 \\
  1 & \beta_0 & \dots & \beta_0^{a-1} \\
  1 & \beta_0^2 & \dots & \beta_0^{2(a-1)} \\
  \vdots & \vdots & & \vdots \\
  1 & \beta_0^{\delta} & \dots & \beta_0^{\delta(a-1)}
\end{pmatrix}.
\]
As before, if $\mathset{1,\beta_0,\beta_0^2,\dots,\beta_0^\delta}\neq
\mathset{1,\beta_i,\beta_i^2,\dots,\beta_i^\delta}$, then by the fact
that $r\geq 3$, we would have a contradiction to the rank equality
above. If follows that
\begin{equation}
  \label{eq:seqbeta2}
\mathset{1,\beta_0,\beta_0^2,\dots,\beta_0^\delta}=
\mathset{1,\beta_i,\beta_i^2,\dots,\beta_i^\delta},
\end{equation}
for all $i\in\Z_m$.

The combination of~\eqref{eq:seqbeta} and~\eqref{eq:seqbeta2} implies that
$\beta_0^\delta=\beta_i^\delta$ for all $i\in\Z_m$. We observe that
\begin{align*}
  \beta_0^\delta-1 & = (1+\beta_0+\dots+\beta_0^{\delta-1})(\beta_0-1), \\
  \beta_i^\delta-1 & = (1+\beta_i+\dots+\beta_i^{\delta-1})(\beta_i-1). \\
\end{align*}
By~\eqref{eq:seqbeta},
\[ \frac{\beta_0^\delta-1}{\beta_i^\delta-1}=\frac{\beta_0-1}{\beta_i-1}.\]
But now, since $\beta_0^\delta=\beta_i^\delta$, we conclude that
\[ \beta_i=\beta_0,\]
for all $i\in\Z_m$.

Now that we know that $\beta_0=\beta_1=\dots=\beta_{m-1}$, we can
write $H_\ell$ as
\[
  H_\ell=
\parenv{\begin{array}{cccc|cccc|c|ccc}
\vv{\beta_0^{\tau_0}}&\vv{0}&\cdots &\vv{0}&\vv{\beta_0^{\tau_0+1}}&\vv{0}&\cdots &\vv{0}&\cdots&\vv{\beta_0^{\tau_0+a-1}}&\cdots &\vv{0}\\
\vv{0}&\vv{\beta_0^{\tau_1}}&\cdots &\vv{0}&\vv{0}&\vv{\beta_0^{\tau_1+1}}&\cdots &\vv{0}&\cdots&\vv{0}&\cdots &\vv{0}\\
\vdots&\vdots&\ddots &\vdots&\vdots&\vdots&\ddots &\vdots& &\vdots&\ddots &\vdots\\
\vv{0}&\vv{0}&\cdots &\vv{\beta_0^{\tau_{m-1}}}&\vv{0}&\vv{0}&\cdots &\vv{\beta_0^{\tau_{m-1}+1}}&\cdots&\vv{0}&\cdots &\vv{\beta_0^{\tau_{m-1}+a-1}}\\
\lambda^{\zeta_0}&\lambda^{\zeta_1}&\cdots &\lambda^{\zeta_{m-1}}&\lambda^{\zeta_0}&\lambda^{\zeta_1}&\cdots &\lambda^{\zeta_{m-1}}&\cdots&\lambda^{\zeta_0}&\cdots &\lambda^{\zeta_{m-1}}\\
\beta_0^{\tau_0\delta}&\beta_0^{\tau_1\delta}&\cdots &\beta_0^{\tau_{m-1}\delta}&\beta_0^{(\tau_0+1)\delta}
&\beta_0^{(\tau_1+1)\delta}&\cdots &\beta_0^{(\tau_{m-1}+1)\delta}&\cdots&\beta_0^{\delta (\tau_0+a-1)}&\cdots &\beta_0^{\delta(\tau_{m-1}+a-1)}\\
\end{array}}.
\]
Looking at the columns of $H_\ell$ that correspond to $G_j$,
$j\in\Z_m$, once again we observe that the non-zero rows are
equivalent to a (transposed) Vandermonde matrix
\[ \begin{pmatrix}
  1 & 1 & \dots & 1 \\
  1 & \beta_0 & \dots & \beta_0^{a-1} \\
  1 & \beta_0^2 & \dots & \beta_0^{2(a-1)} \\
  \vdots & \vdots & & \vdots \\
  1 & \beta_0^{\delta} & \dots & \beta_0^{\delta(a-1)}
\end{pmatrix}.
\]
Hence, these rows are linearly independent. Let us now add a
cyclically shifted version of the last row of $H_\ell$, to obtain yet
another parity-check matrix for the code,
\[
  H^*_\ell=
\parenv{\begin{array}{cccc|cccc|c|ccc}
\vv{\beta_0^{\tau_0}}&\vv{0}&\cdots &\vv{0}&\vv{\beta_0^{\tau_0+1}}&\vv{0}&\cdots &\vv{0}&\cdots&\vv{\beta_0^{\tau_0+a-1}}&\cdots &\vv{0}\\
\vv{0}&\vv{\beta_0^{\tau_1}}&\cdots &\vv{0}&\vv{0}&\vv{\beta_0^{\tau_1+1}}&\cdots &\vv{0}&\cdots&\vv{0}&\cdots &\vv{0}\\
\vdots&\vdots&\ddots &\vdots&\vdots&\vdots&\ddots &\vdots& &\vdots&\ddots &\vdots\\
\vv{0}&\vv{0}&\cdots &\vv{\beta_0^{\tau_{m-1}}}&\vv{0}&\vv{0}&\cdots &\vv{\beta_0^{\tau_{m-1}+1}}&\cdots&\vv{0}&\cdots &\vv{\beta_0^{\tau_{m-1}+a-1}}\\
\lambda^{\zeta_0}&\lambda^{\zeta_1}&\cdots &\lambda^{\zeta_{m-1}}&\lambda^{\zeta_0}&\lambda^{\zeta_1}&\cdots &\lambda^{\zeta_{m-1}}&\cdots&\lambda^{\zeta_0}&\cdots &\lambda^{\zeta_{m-1}}\\
\beta_0^{\tau_0\delta}&\beta_0^{\tau_1\delta}&\cdots &\beta_0^{\tau_{m-1}\delta}&\beta_0^{(\tau_0+1)\delta}
&\beta_0^{(\tau_1+1)\delta}&\cdots &\beta_0^{(\tau_{m-1}+1)\delta}&\cdots&\beta_0^{\delta (\tau_0+a-1)}&\cdots &\beta_0^{\delta(\tau_{m-1}+a-1)}\\ \hline
\beta_0^{\tau_1\delta}&\beta_0^{\tau_2\delta}&\cdots &\beta_0^{(\tau_0+1)\delta}&\beta_0^{(\tau_1+1)\delta}&\beta_0^{(\tau_2+1)\delta}&\cdots &\beta_0^{(\tau_0+2)\delta}&\cdots&\beta_0^{\delta (\tau_1+a-1)}&\cdots &\beta_0^{\tau_0 \delta}\\
\end{array}}.
\]
The added row is a linear combination of the original rows of
$H_\ell$. Assume $j\in\Z_m$, $j\neq m-1$. If we look at the columns of $G_j$
in $H^*_\ell$ we see that
\[ \parenv{\beta_0^{\tau_{j+1} \delta},\beta_0^{(\tau_{j+1}+1)\delta}, \dots, \beta_0^{(\tau_{j+1}+a-1)\delta}} = \beta^{(\tau_{j+1}-\tau_j)\delta}\parenv{\beta_0^{\tau_{j} \delta},\beta_0^{(\tau_{j}+1)\delta}, \dots, \beta_0^{(\tau_{j}+a-1)\delta}}.\]
Since the non-zero rows in the columns of $G_j$ are linearly
independent, this linear combination is unique. Similarly, for $j=m-1$ we get
\[ \parenv{\beta_0^{(\tau_{0}+1) \delta},\beta_0^{(\tau_{0}+2)\delta}, \dots, \beta_0^{\tau_{0}\delta}} = \beta^{(\tau_{0}-\tau_{m-1})\delta+\delta}\parenv{\beta_0^{\tau_{m-1} \delta},\beta_0^{(\tau_{m-1}+1)\delta}, \dots, \beta_0^{(\tau_{m-1}+a-1)\delta}},\]
which is again unique. However, all these linear combinations must
coincide simultaneously when viewing the entire $H^*_\ell$, and so
\[ \beta_0^{(\tau_1-\tau_0)\delta}=\beta_0^{(\tau_2-\tau_1)\delta}= \dots = \beta_0^{(\tau_{m-1}-\tau_{m-2})\delta}=\beta_0^{(\tau_0-\tau_{m-1})\delta+\delta}.\]
Multiplying all of them together we get
\[ \parenv{\beta_0^{(\tau_1-\tau_0)\delta}}^m = \beta_0^{(\tau_1-\tau_0)\delta}\cdot\ldots\cdot\beta_0^{(\tau_{m-1}-\tau_{m-2})\delta}\cdot\beta_0^{(\tau_0-\tau_{m-1})\delta+\delta} = \beta_0^\delta.\]
However, this means that
\[ \delta m (\tau_1-\tau_0) \equiv \delta \pmod{a},\]
which completes the proof.
\end{IEEEproof}

While the last theorem shows us a sufficient condition under which
the known code of Construction~\ref{cons_MR_known} may be permuted to
a cyclic code, the next theorem shows us that for almost all cases, this
condition is in fact necessary. First, we bring a technical proposition.

\begin{proposition}
  \label{prop_tau}
  Let $a$, $r$, $\delta$ be positive integers with $a=r+\delta-1$, and $\tau,\tau'\in \euler{a}$.  If
  \begin{equation}
    \label{eqn_cond_tau}
    \mathset{i\tau \bmod a~:~1\leq i\leq\delta-1}\subseteq \mathset{i\tau' \bmod a ~:~1\leq i\leq \delta},
  \end{equation}
  and one of the following conditions holds,
  \begin{enumerate}
  \item\label{it:tau1}
    $\delta\geq 4$ and $r\geq 5$
  \item\label{it:tau2}
    $\delta=3$ and $r\geq 4$
  \item\label{it:tau3}
    $\delta=2$ and $r\geq 3$ is odd
  \end{enumerate}
  then we have $\tau=\tau'$.
\end{proposition}

\begin{IEEEproof}
  For Case~\ref{it:tau1}, since $\tau, \tau'\in \euler{a}$ there
  exists an $\ell\in \euler{a}$ such that $\tau\equiv \ell\tau'\pmod
  a$.  By~\eqref{eqn_cond_tau}, we have
  \[
    \mathset{(i+1)\tau\bmod a~:~ 1\leq i\leq \delta-1}\subseteq \mathset{(i\tau'+\tau)\bmod a~:~1\leq i\leq \delta}=\mathset{(i+\ell)\tau'\bmod a~:~1\leq i\leq\delta}.
  \]
  Obviously,
  \[\abs{\mathset{(i+1)\tau\bmod a~:~1\leq i\leq \delta-1}\cap\mathset{i\tau\bmod a~:~1\leq i\leq \delta-1}}\geq \delta-2,\]
  and so,
  \[\abs{\mathset{i\tau'\bmod a~:~1\leq i\leq \delta}\cap \mathset{(i+\ell)\tau'\bmod a~:~1\leq i\leq \delta}}\geq \delta-2,\]
  and since $\tau'\in \euler{a}$,
  \[\abs{\mathset{i\bmod a~:~1\leq i\leq \delta}\cap \mathset{(i+\ell)\bmod a~:~1\leq i\leq \delta}}\geq \delta-2.\]
  Since $\delta-1\geq 3$ and $a\geq \delta+4$, we must have $\ell\in
  \mathset{1,2}$. It remains to show that $\ell\neq 2$. Assume to the
  contrary that $\ell=2$. Similarly, by \eqref{eqn_cond_tau}, we have
  \[
  \mathset{(i+2)\tau\bmod a~:~1\leq i\leq \delta-1}\subseteq \mathset{(i\tau'+2\tau)\bmod a~:~1\leq i\leq \delta}=\mathset{(i+4)\tau'\bmod a~:~1\leq i\leq \delta}.
  \]
  Again,
  \[\abs{\mathset{(i+2)\tau\bmod a~:~1\leq i\leq \delta-1}\cap \mathset{i\tau\bmod a~:~1\leq i\leq \delta-1}}\geq \delta-3,\]
  which implies that,
  \[\abs{\mathset{(i+4)\tau'\bmod a~:~1\leq i\leq \delta}\cap \mathset{i\tau'\bmod a~:~1\leq i\leq \delta}}\geq \delta-3,\]
and since $\tau'\in \euler{a}$,
\[\abs{\mathset{(i+4)\bmod a~:~1\leq i\leq \delta}\cap \mathset{i\bmod a~:~1\leq i\leq \delta}}\geq \delta-3.\]
However, $a\geq \delta+4$ implies that
\[\abs{\mathset{(i+4)\bmod a~:~1\leq i\leq \delta}\cap \mathset{i\bmod a~:~1\leq i\leq \delta}}\leq \delta-4,\]
which is a contradiction. Thus, we have $\ell=1$ and $\tau=\tau'$.

For Case~\ref{it:tau2}, by~\eqref{eqn_cond_tau}, we have
\[\mathset{\tau,2\tau \bmod a}\subseteq \mathset{\tau',2\tau'\bmod a,3\tau'\bmod a}.\]
If $\tau\equiv 2\tau'\pmod{a}$ then we have $2\tau\equiv
4\tau'\pmod{a}$, hence $4\tau' \bmod a\in \mathset{\tau',2\tau'\bmod a,3\tau'\bmod a}$. However, this is impossible since
$\tau'\in \euler{a}$ and $a\geq 6$. Similarly, if $\tau\equiv
3\tau'\pmod{a}$ then we have $2\tau\equiv 6\tau'\pmod{a}$, hence
$6\tau'\bmod a\in \mathset{\tau',2\tau'\bmod a,3\tau'\bmod
  a}$. Again, this is also impossible since $\tau'\in \euler{a}$ and
$a\geq 6$. Thus, we have $\tau=\tau'$.

Finally, for Case~\ref{it:tau3}, by~\eqref{eqn_cond_tau} we have
$\mathset{\tau}\subseteq \mathset{\tau',2\tau'\bmod a}$. Obviously
$\tau\equiv 2\tau'\pmod{a}$ is impossible since $2|a$ and
$a\geq 3$. Thus, $\tau=\tau'$.
\end{IEEEproof}

\begin{theorem}
  \label{th:cyclic}
  Assume the notation and setting of Construction~\ref{cons_MR_known}.
  Let $\cC$ be the resulting code. Denote $k=\dim(\cC)=ur+v$ with
  $0<v\leq r$ and $u\geq 2(r-v+1)$. Additionally, let $a=4$ or
  $\gcd(a,\varphi(a))=1$. Furthermore, assume that $a=q^{b'}-1|q^b-1=n$,
  and that one of the following holds:
  \begin{enumerate}
  \item
    $\delta\geq 4$ and $r\geq 5$
  \item
    $\delta=3$ and $r\geq 4$
  \item
    $\delta=2$ and $r\geq 3$ is odd
  \end{enumerate}
  Then there exists a permutation $\ell\in\bS_n$ such that $\cC_\ell$
  is cyclic only if $\gcd(m,a)\mid \delta$.
\end{theorem}

\begin{IEEEproof}
  Assume $\ell\in\bS_n$ is a
  permutation such that $\cC_\ell$ is cyclic. By
  Construction~\ref{cons_MR_known}, we have that $G_i\eqdef
  \angleenv{m}+i$, $i\in[m]$, are exactly the repair sets of
  $\cC$. Thus, the image sets $\ell(G_i)\eqdef \mathset{\ell(x)~:~x\in
    G_i}$ are exactly the repair sets of $\cC_{\ell}$. Note that $\cC$
  is an optimal LRC, which means that $\cC_{\ell}$ is also optimal. By
  Corollary~\ref{corollary_repair_set}, we have
  \[\mathset{\ell(G_i)~:~ i\in [m]}=\mathset{G_i~:~ i\in [m] }.\]
  Thus, there exists a sequence of permutations, $\ell_i$ over $G_i$,
  for all $i\in [m]$, and $z_i\in \Z$, such that for all $x\in G_i$,
  \begin{equation}
    \label{eq:permell}
    \ell(x)=\parenv{\ell_i(x)+z_i} \bmod n,
  \end{equation}
  which also implies that $\ell(G_i)=G_{i+z_i}$, and
  $(z_0,\dots,z_{m-1})$ is a permutation of $[m]$. By assumption,
  $\cC_{\ell}$ is cyclic. Hence, $\cC_{\ell}|_{G_i}$ is also cyclic,
  for each $i\in [m]$.

  By Construction~\ref{cons_MR_known}, any punctured code,
  $\cC|_{G_i}$, $i\in [m]$, is a subcode of the code with the
  $(\delta-1)\times a$ parity-check matrix
  $(\vv{1},\vv{\beta},\dots,\vv{\beta^{a-1}})$. Recall that $\cC$ is
  an optimal LRC. Hence, by Theorem~\ref{lemma_repair_sets}, we have
  that this punctured code, $\cC|_{G_i}$, is an
  $[a=r+\delta-1,r,\delta]_q$ MDS code. This implies that its
  parity-check matrix is \emph{exactly}
  $(\vv{1},\vv{\beta},\dots,\vv{\beta^{a-1}})$. Since this matrix
  clearly does not depend on $i$, we have $\cC|_{G_i}=\cC|_{G_j}$, for
  all $i,j\in [m]$. Additionally, since $\beta^a=1$, all the punctured
  codes $\cC|_{G_i}$ are cyclic.

  By Corollary \ref{corollary_ parity}, we have that
  $\cC_{\ell}|_{G_i}=\cC_{\ell}|_{G_j}$ for all $i,j\in [m]$, and are
  all cyclic codes. Thus, a parity-check matrix of $\cC_{\ell}|_{G_i}$
  may be given by
  \begin{equation}
    \label{eq:hclgi}
    H(\cC_\ell|_{G_i})=
    (\vv{\beta^{\tau_{i,0}}}, \vv{\beta^{\tau_{i,1}}}, \cdots ,\vv{\beta^{\tau_{i,a-1}}}).
  \end{equation}
 We now have that
  $\ell_0$ maps the cyclic code $\cC|_{G_0}$ into a cyclic code
  $\cC_\ell|_{G_0}$, where we view these codes as indexed by $\Z_a$.
  Then, by Lemma~\ref{lemma_multipliers}, we can find a multiplier permutation
  from $\Upsilon(a)$ that also maps $\cC|_{G_0}$ to $\cC_\ell|_{G_0}$. More
  concretely, there exists $\tau'\in\euler{a}$, with which we define a permutation
  \[ \ell'(xm+i) = (x m \tau' + i) \pmod{n},\]
  for all $x\in[a]$ and $i\in[m]$. For this permutation we have
  $\cC_{\ell}|_{G_i}=\cC_{\ell'}|_{G_i}$ for all $i\in [m]$. Now, a parity-check
  matrix for $\cC_{\ell'}|_{G_i}$ may be given by
  \[ H(\cC_{\ell'}|_{G_i})=
  (\vv{1}, \vv{\beta^{\tau'}}, \cdots ,\vv{\beta^{\tau'(a-1)}}),
  \]
  and it must be row-equivalent to $H(\cC_{\ell}|_{G_i})$
  from~\eqref{eq:hclgi}.

  We switch our view from the punctured codes to the shortened codes
  of $\cC$. As above, the following shortened codes are all equal,
  $\cC|^{G_i}=\cC|^{G_0}$, for all $i\in[m]$. A parity-check matrix
  for them may be written as
  \begin{equation*}
    \begin{pmatrix}
      1 & 1 & \cdots & 1 \\
      \vv{1} & \vv{\beta} & \cdots & \vv{\beta^{a-1}} \\
      1 & \beta^{\delta} & \cdots & \beta^{\delta(a-1)}
    \end{pmatrix}
    =
    \begin{pmatrix}
      1&1&\cdots&1\\
      1 &\beta &\cdots &\beta^{a-1}\\
      1 &\beta^2 &\cdots &\beta^{2(a-1)}\\
      \vdots &\vdots &  &\vdots\\
      1 &\beta^{\delta-1} &\cdots &\beta^{(\delta-1)(a-1)}\\
      1 &\beta^{\delta} &\cdots &\beta^{(\delta)(a-1)}\\
    \end{pmatrix}.
  \end{equation*}
  By the multiplicative order of $\beta$, the codes $\cC|^{G_i}$ are
  all cyclic.  By Corollary~\ref{corollary_ parity},
  $\cC_{\ell}|^{G_i}=\cC_{\ell}|^{G_0}$, for all $i\in[m]$, and a
  parity-check matrix for them may be given by
  \begin{equation}\label{eqn_short_H_ell_i}
    H(\cC_\ell|^{G_i})=\begin{pmatrix}
    1&1&\cdots&1\\
    \vv{\beta^{\tau_{i,0}}} & \vv{\beta^{\tau_{i,1}}} &\cdots &\vv{\beta^{\tau_{i,a-1}}}\\
    \beta^{\delta\tau_{i,0}} &\beta^{\delta\tau_{i,1}} &\cdots &\beta^{\delta\tau_{i,a-1}}\\
    \end{pmatrix},
  \end{equation}
  where $\tau_{i,j}$, $i\in[m]$, $j\in[a]$, are the same as those
  in~\eqref{eq:hclgi}.  Once again, $\cC_\ell|^{G_i}$ are all
  cyclic. Hence, by Lemma~\ref{lemma_multipliers} we can find a
  multiplier permutation from $\Upsilon(a)$ that also maps
  $\cC|^{G_0}$ to $\cC_\ell|^{G_0}$. Namely, there exists
  $\tau''\in\euler{a}$, with which we define
  \[ \ell''(xm+i)=(xm\tau''+i)\pmod{n},\]
  for all $x\in[a]$ and $i\in[m]$, such that
  $\cC_{\ell}|^{G_i}=\cC_{\ell''}|^{G_i}$, for all $i\in[m]$. A parity-check
  matrix for $\cC_{\ell''}|^{G_i}$ may be given by
  \[ H(\cC_{\ell''}|^{G_i})=
  \begin{pmatrix}
    1&1&\cdots&1\\
    \vv{1} & \vv{\beta^{\tau''}} &\cdots &\vv{\beta^{\tau''(a-1)}}\\
    1 &\beta^{\delta\tau''} &\cdots &\beta^{\delta\tau''(a-1)}\\
    \end{pmatrix},
  \]
  and it must be row-equivalent to $H(\cC_{\ell}|^{G_i})$
  from~\eqref{eqn_short_H_ell_i}.

  By the properties of Vandermonde matrices we have for all $i\in[m]$,
  \begin{align*}
      \delta+1
      &=
      \rank\begin{pmatrix}
      1&1&\cdots&1\\
      \vv{\beta^{\tau_{i,0}}} & \vv{\beta^{\tau_{i,1}}} &\cdots & \vv{\beta^{\tau_{i,a-1}}}\\
      \beta^{\delta\tau_{i,0}} &\beta^{\delta\tau_{i,1}} &\cdots &\beta^{\delta\tau_{i,a-1}}\\
      \end{pmatrix}
      =
      \rank\begin{pmatrix}
      1&1&\cdots&1\\
      \vv{\beta^{\tau_{i,0}}} & \vv{\beta^{\tau_{i,1}}} &\cdots & \vv{\beta^{\tau_{i,a-1}}}\\
      \vv{\beta^{\tau_{i,0}}} & \vv{\beta^{\tau_{i,1}}} &\cdots & \vv{\beta^{\tau_{i,a-1}}}\\
      \beta^{\delta\tau_{i,0}} &\beta^{\delta\tau_{i,1}} &\cdots &\beta^{\delta\tau_{i,a-1}}\\
      \end{pmatrix}\\
      &=
      \rank\begin{pmatrix}
      1&1&\cdots&1\\
      \vv{1} & \vv{\beta^{\tau'}} &\cdots & \vv{\beta^{\tau'(a-1)}}\\
      \vv{1} & \vv{\beta^{\tau''}} &\cdots & \vv{\beta^{\tau''(a-1)}}\\
      1 &\beta^{\delta\tau''} &\cdots &\beta^{\delta\tau''(a-1)}\\
      \end{pmatrix},
  \end{align*}
  where the last equality holds by the row equivalence of
  $H(\cC_\ell|_{G_i})$ and $H(\cC_{\ell'}|_{G_i})$, as well as the row
  equivalence of $H(\cC_\ell|^{G_i})$ and
  $H(\cC_{\ell''}|^{G_i})$. Since $r\geq 3$, the above equality
  implies
  \[\mathset{\beta^{j\tau'}~:~1\leq j\leq \delta-1}\subseteq
  \mathset{\beta^{j\tau''}~:~0\leq j\leq \delta}.\]
  By construction, the multiplicative order of $\beta$ is
  $o(\beta)=a$, and so
  \[ o(\beta^{\tau'})=\frac{o(\beta)}{\gcd(\tau',o(\beta))}=\frac{a}{\gcd(\tau',a)}=a,\]
  where the last equality follows from the fact that
  $\tau'\in\euler{a}$. Since $a=r+\delta-1$,
  \[1\not\in \mathset{\beta^{j\tau'}~:~1\leq j\leq \delta-1}.\]
  Thus,
  \[
  \mathset{\beta^{j\tau'}~:~1\leq j\leq \delta-1}\subseteq \mathset{\beta^{j\tau''}~:~1\leq j\leq \delta}.
  \]
  Since $o(\beta)=a$, we have
  \[\mathset{j\tau'\bmod a~:~1\leq j\leq \delta-1}\subseteq \mathset{j\tau''\bmod a~:~1\leq j\leq \delta}.\]
  Then, by Proposition~\ref{prop_tau}, we have $\tau'=\tau''$.

  Denote $\gamma\eqdef\beta^{\tau'}=\beta^{\tau''}$. Thus, $(\vv{1},
  \vv{\gamma}, \cdots, \vv{\gamma^{a-1}})=(\vv{1},
  \vv{\beta^{\tau'}}, \cdots, \vv{\beta^{\tau'(a-1)}})$.  We now know
  that the following two matrices are row equivalent,
  \begin{equation}
    \label{eq:betagamma}
  \begin{pmatrix}
    1&1&\cdots&1\\
    \vv{1} & \vv{\gamma} &\cdots & \vv{\gamma^{a-1}}\\
    1 &\gamma^{\delta} &\cdots &\gamma^{\delta(a-1)}\\
  \end{pmatrix} \quad\text{and}\quad
  \begin{pmatrix}
    1&1&\cdots&1\\
    \vv{1} & \vv{\gamma} &\cdots & \vv{\gamma^{a-1}}\\
    \beta^{\delta\tau_{i,0}} &\beta^{\delta\tau_{i,1}} &\cdots &\beta^{\delta\tau_{i,a-1}}\\
  \end{pmatrix},
  \end{equation}
  for all $i\in[m]$. Recall that $\beta$ and $\gamma$ have the same
  order, $o(\beta)=o(\gamma)a=q^{b'}-1$, i.e., the entries of the matrices
  in~\eqref{eq:betagamma} belong to the field $\F_{q^{b'}}$.
  Hence,
  $(\beta^{\delta\tau_{i,0}},\beta^{\delta\tau_{i,1}},\cdots,\beta^{\delta\tau_{i,a-1}})$
  can be represented as a linear combination
  \begin{equation}\label{eqn_rep}
    (\beta^{\delta\tau_{i,0}},\beta^{\delta\tau_{i,1}},\cdots,\beta^{\delta\tau_{i,a-1}})
    =\sum_{s=0}^{\delta}\eta_{i,s} (1,\gamma^{s},\cdots,\gamma^{s(a-1)})
  \end{equation}
  where $\eta_{i,s}\in \F_{q^{b'}}\subseteq \F_{q^b}$ for all
  $i\in[m]$, $s\in[\delta+1]$. We also highlight the fact that
  $\eta_{i,\delta}\neq 0$ for all $i\in[m]$, for otherwise we would
  have that the matrix on the right has rank $\delta$ whereas the one
  on the left has rank $\delta+1$. For convenience, let us define
  $\xi_{i,j}\eqdef \eta_{i,\delta}\gamma^{\delta j}+\eta_{i,0}$, where
  $i\in[m]$, $j\in[a]$.

  After focusing on shortened and punctured codes, let us look again
  at the entire code. If we permute the columns of the parity-check
  matrix of $\cC$ using $\ell$, we arrive at the following
  parity-check matrix for $\cC_\ell$, to \eqref{eqn_short_H_ell_i},
  \begin{equation*}
    H_{\ell}=\parenv{\begin{array}{cccc|cccc|c|cccc}
        \vv{\beta^{\tau_{0,0}}}&0&\cdots&0 &\vv{\beta^{\tau_{0,1}}}&0&\cdots &0&\cdots&\vv{\beta^{\tau_{0,a-1}}}&0&\cdots &0\\
        0&\vv{\beta^{\tau_{1,0}}}&\cdots &0&0&\vv{\beta^{\tau_{1,1}}}&\cdots &0&\cdots&0&\vv{\beta^{\tau_{1,a-1}}}&\cdots &0\\
        \vdots&\vdots& &\vdots&\vdots&\vdots& &\vdots& &\vdots&\vdots& &\vdots\\
        0&0&\cdots &\vv{\beta^{\tau_{m-1,0}}}&0&0&\cdots &\vv{\beta^{\tau_{m-1,1}}}&\cdots&0&0&\cdots &\vv{\beta^{\tau_{m-1,a-1}}}\\
        \lambda^{z_0}&\lambda^{z_1}&\cdots &\lambda^{z_{m-1}}&\lambda^{z_0}&\lambda^{z_1}&\cdots &\lambda^{z_{m-1}}&\cdots&\lambda^{z_0}&\lambda^{z_1}&\cdots &\lambda^{z_{m-1}}\\
        \beta^{\delta\tau_{0,0}}&\beta^{\delta\tau_{1,0}}&\cdots &\beta^{\delta\tau_{m-1,0}}&\beta^{\delta\tau_{0,1}}&\beta^{\delta\tau_{1,1}}
        &\cdots &\beta^{\delta\tau_{m-1,1}}&\cdots&\beta^{\delta\tau_{0,a-1}}&\beta^{\delta\tau_{1,a-1}}&\cdots &\beta^{\delta\tau_{m-1,a-1}}\\
    \end{array}},
  \end{equation*}
  where $z_i$, $i\in[m]$ are the same as in~\eqref{eq:permell}, and
  $\tau_{i,j}$, $i\in[m]$, $j\in[a]$, are the same as
  in~\eqref{eq:hclgi}. By~\eqref{eqn_rep} and the equivalence of
  $H(\cC_\ell|_{G_i})$ and $H(\cC_{\ell'}|_{G_i})$, the matrix $H_\ell$
  is row equivalent with
  \begin{equation*}
    H'=\parenv{\begin{array}{cccc|cccc|c|cccc}
        \vv{1}&0&\cdots&0 &\vv{\gamma}&0&\cdots &0&\cdots&\vv{\gamma^{a-1}}&0&\cdots &0\\
        0&\vv{1}&\cdots &0&0&\vv{\gamma}&\cdots &0&\cdots&0&\vv{\gamma^{a-1}}&\cdots &0\\
        \vdots&\vdots& &\vdots&\vdots&\vdots& &\vdots& &\vdots&\vdots& &\vdots\\
        0&0&\cdots &\vv{1}&0&0&\cdots &\vv{\gamma}&\cdots&0&0&\cdots &\vv{\gamma^{a-1}}\\
        \lambda^{z_0}&\lambda^{z_1}&\cdots &\lambda^{z_{m-1}}&\lambda^{z_0}&\lambda^{z_1}&\cdots &\lambda^{z_{m-1}}&\cdots&\lambda^{z_0}&\lambda^{z_1}&\cdots &\lambda^{z_{m-1}}\\
        \xi_{0,0}& \xi_{1,0} &\cdots & \xi_{m-1,0} & \xi_{0,1} & \xi_{1,1}
        &\cdots &\xi_{m-1,1}&\cdots&\xi_{0,a-1} & \xi_{1,a-1}
        &\cdots & \xi_{m-1,a-1}\\
    \end{array}}.
  \end{equation*}
  Since $H'$ is also a parity-check matrix for $\cC_\ell$, which is a cyclic
  code, adding a dependent row to $H'$ which is a cyclic shift of another row,
  does not change the code. Hence, we look at the following parity-check
  matrix for $\cC_\ell$,
  \begin{equation*}
    H''=\parenv{\begin{array}{cccc|cccc|c|cccc}
        \vv{1}&0&\cdots&0 &\vv{\gamma}&0&\cdots &0&\cdots&\vv{\gamma^{a-1}}&0&\cdots &0\\
        0&\vv{1}&\cdots &0&0&\vv{\gamma}&\cdots &0&\cdots&0&\vv{\gamma^{a-1}}&\cdots &0\\
        \vdots&\vdots& &\vdots&\vdots&\vdots& &\vdots& &\vdots&\vdots& &\vdots\\
        0&0&\cdots &\vv{1}&0&0&\cdots &\vv{\gamma}&\cdots&0&0&\cdots &\vv{\gamma^{a-1}}\\
        \lambda^{z_0}&\lambda^{z_1}&\cdots &\lambda^{z_{m-1}}&\lambda^{z_0}&\lambda^{z_1}&\cdots &\lambda^{z_{m-1}}&\cdots&\lambda^{z_0}&\lambda^{z_1}&\cdots &\lambda^{z_{m-1}}\\
        \xi_{0,0}& \xi_{1,0} &\cdots & \xi_{m-1,0} & \xi_{0,1} & \xi_{1,1}
        &\cdots &\xi_{m-1,1}&\cdots&\xi_{0,a-1} & \xi_{1,a-1}
        &\cdots & \xi_{m-1,a-1}\\ \hline
        \xi_{1,0}& \xi_{2,0} &\cdots & \xi_{0,1} & \xi_{1,1} & \xi_{2,1}
        &\cdots &\xi_{0,2}&\cdots&\xi_{1,a-1} & \xi_{2,a-1}
        &\cdots & \xi_{0,0}\\
    \end{array}}.
  \end{equation*}

  Let us now denote by $h$ the bottom row of $H''$, and by
  $h_{-2},h_{-1}$ the bottom two rows of $H'$. We recall that
  $\xi_{i,j}\eqdef \eta_{i,\delta}\gamma^{\delta j}+\eta_{i,0}$, and hence,
  \begin{align*}
    h|_{G_i} & = ( \eta_{i+1,\delta}\gamma^{0}+\eta_{i+1,0}, \eta_{i+1,\delta}\gamma^\delta+\eta_{i+1,0},\dots,\eta_{i+1,\delta}\gamma^{\delta(a-1)}+\eta_{i+1,0}), & i&\in[m-1]\\
    h|_{G_{m-1}} & = ( \eta_{0,\delta}\gamma^{\delta}+\eta_{0,0}, \eta_{0,\delta}\gamma^{2\delta}+\eta_{0,0},\dots,\eta_{0,\delta}\gamma^{0}+\eta_{0,0}),\\
    h_{-1}|_{G_i} & = ( \eta_{i,\delta}\gamma^{0}+\eta_{i,0}, \eta_{i,\delta}\gamma^\delta+\eta_{i,0},\dots,\eta_{i,\delta}\gamma^{\delta(a-1)}+\eta_{i,0}), & i&\in[m]\\
    h_{-2}|_{G_i} & = \lambda^{z_i}(1,1,\dots,1). & i&\in[m]
  \end{align*}
  We now observe that the last row of $H''|_{G_i}$ may be shown as a
  linear combination of the preceding two rows. More precisely, for
  all $i\in[m]$,
  \[ h|_{G_i} = \theta_{i,1}h_{-1}|_{G_i} + \theta_{i,2}h_{-2}|_{G_i},\]
  where
  \begin{align}
    \label{eq:thetadef}
    \theta_{i,1}&=\begin{cases}
    \frac{\eta_{i+1,\delta}}{\eta_{i,\delta}}& i\in[m-1],\\
    \frac{\eta_{0,\delta}}{\eta_{m-1,\delta}}\gamma^{\delta}& i=m-1,\\
    \end{cases} \\
    \theta_{i,2}&=\begin{cases}
    \frac{\eta_{i+1,0}-\theta_{1,i}\eta_{i,0}}{\lambda^{z_i}}& i\in[m-1],\\
    \frac{\eta_{0,0}-\theta_{m-1,1}\eta_{m-1,0}}{\lambda^{z_{m-1}}}& i=m-1.
    \end{cases}\nonumber
  \end{align}
  Since $H''$ is row equivalent with $H'$, and
  $\rank(H'|_{G_i})=\delta+1$ (i.e., full rank), the linear
  combination above is the \emph{unique} linear combination of the
  rows of $H'|_{G_i}$ that gives $h|_{G_i}$. This linear combination
  does not use the first $\delta-1$ rows of $H'|_{G_i}$.

  Looking at the entire matrix (instead of focusing on the projections
  onto $G_i$), once again, since $H''$ is row equivalent with $H'$,
  $h$ must be linear combination of the rows of $H'$. Since in each
  projection onto $G_i$ there is a unique linear combination, all these
  must simultaneously agree. In particular, this means
  \[ \theta_{0,1} = \theta_{1,1} = \theta_{2,1} = \dots = \theta_{m-1,1}.\]
  We recall that $0\neq \eta_{i,\delta}\in\F_{q^{b'}}$ for all $i\in[m]$, and
  $\gamma\in\F_{q^{b'}}$ is primitive. Thus, we may write
  \begin{equation}
    \label{eq:theta1}
    \theta_{0,1} = \theta_{1,1} = \theta_{2,1} = \dots = \theta_{m-1,1}=\gamma^j,
  \end{equation}
  for some integer $j$. Also, by~\eqref{eq:thetadef},
  \begin{equation}
    \label{eq:theta2}
    \frac{\eta_{1,\delta}}{\eta_{0,\delta}} = \frac{\eta_{2,\delta}}{\eta_{1,\delta}} = \frac{\eta_{3,\delta}}{\eta_{2,\delta}}=\dots=\frac{\eta_{m-1,\delta}}{\eta_{m-2,\delta}}=\gamma^\delta \frac{\eta_{0,\delta}}{\eta_{m-1,\delta}}.
  \end{equation}
  Now, combining~\eqref{eq:theta1} and~\eqref{eq:theta2} we get
  \[ \gamma^{jm} = \prod_{i\in[m]}\theta_{i,1} = \gamma^{\delta}.\]
  Thus,
  \[ jm \equiv \delta \pmod{a}.\]
  This, in turn, implies that $\gcd(m,a)\mid \delta$, as
  we wanted to prove.
\end{IEEEproof}

  To conclude this section, we make use of Theorem~\ref{th:cyclic} in
  order to show that Construction~\ref{cons_CPMDS} may produce cyclic
  MR codes with new parameters. Namely, in certain case, the
  construction of~\cite{gopi2020maximally}, which produces codes with
  the same parameters as our Construction~\ref{cons_CPMDS}, results in
  codes that are neither cyclic, nor can be permuted to become cyclic.

  \begin{example}
    Set $q=3$, $b_1=2$, $b=4$, $r=6$, $\delta=3$, $a=8$, $n=80$, and
    $m=10$. By using Construction \ref{cons_CPMDS}, we may generate a
    cyclic $(n=80,r=6,h=2,\delta=3,q^{b}=3^4)$-MR code, where we note that $\gcd(3,10)=1$. A
    \emph{non-cyclic} MR code with the same parameters may be
    constructed using~\cite{gopi2020maximally}. However, since
    $\gcd(m,a)=\gcd(10,8)=2\nmid 3=\delta$, by
    Theorem~\ref{th:cyclic} this code cannot be permuted to become a
    cyclic code.
  \end{example}

\section{Conclusion}
\label{sec-conclusion}

In this paper, we proved a new lower bound on the field size of
optimal LRCs. As a byproduct, when $r=2$ we were able to prove that
some known code constructions actually have optimal field size (where
we further had to assume that the field size minus $1$ or $2$ is not a
prime power). We then constructed cyclic MR codes. When $r=2$, these
codes also attain the new bound with equality, and therefore have
optimal field size (again, assuming the same number-theoretic
condition). We concluded by showing a known quasi-cyclic MR code, with
the same parameters as our cyclic construction, may sometimes be
permuted to become cyclic, and in other cases it may not.

Many open questions remain. First and foremost, the construction for a
cyclic MR code in this paper only works for the case of two global
parity checks, i.e., $h=2$.  However, in the non-cyclic case, there
are a few known constructions of MR codes with $h\geq 3$. Finding
cyclic MR codes with $h\geq 3$ is still an open question.

As a second open question we mention our lower bound on the field size
of optimal LRCs. We were able to show it is tight only when
$r=2$. Thus, finding out whether it is tight for cases in which $r\geq
3$, or improving it, remains widely open. We leave these questions and
others for future work.

\appendix

In this appendix, we shall prove Theorem~\ref{theorem_repair_set_cyc}.
To this end, we first recall some definitions and lemmas
from~\cite{cai2020optimal}.

Throughout the appendix we shall assume the coordinate of code of
length $n$ are indexed by $\Z_n$, and where operations on coordinates
are required, they shall be made modulo $n$. Let $k=ru+v$ with
$0<v\leq r$. Denote the set of all the possible repair sets for an LRC
$\cC$ with all-symbol $(r,\delta)$-locality as
\[\Gamma\eqdef\mathset{S~:~ S\subseteq\Z_n, \abs{S}\leq r+\delta-1, d(\cC|_{S})\geq \delta}.\]

\begin{lemma}[\cite{cai2019optimal}, Lemma 7]\label{lemma_for_rank_B_i}
  Let $\cC$ be an $[n,k]_q$ linear code with all-symbol
  $(r,\delta)$-locality. If for a subset $\cV\subseteq \Gamma$, and
  for all $S'\in\cV$,
  \[  \abs{S'\cap \parenv{\bigcup_{S\in \cV\setminus\mathset{S'}}S}}\leq
    \abs{S'}-\delta+1,
    \]
  then we have
  \begin{equation*}
    \rank\parenv{\bigcup_{S\in \cV}S}\leq \abs{\bigcup_{S\in \cV}S}-\abs{\cV}(\delta-1).
  \end{equation*}
\end{lemma}

For cyclic LRCs we have the following simple fact.

\begin{lemma}
Let $\cC$ be a cyclic LRC. If $S\in\Gamma$ is a repair set of $\cC$,
then $S+i$ is also a repair set of $\cC$, for all $i\in\Z$.
\end{lemma}
\begin{IEEEproof}
  Since $\cC$ is cyclic, $\cC|_{S}=\cC|_{S+i}$ for any $i\in \Z$. The
  claim follows immediately by definition.
\end{IEEEproof}

We are now ready for the main proof.

\begin{IEEEproof}[Proof of Theorem~\ref{theorem_repair_set_cyc}]
  Assume to the contrary that there exists a repair set $\hS\in \Gamma$
  and $\hatt\in \Z$ such that
  \begin{equation}
    \label{eq:contrar}
    0<\abs{\hS\cap (\hS+\hatt)}<\abs{\hS}.
  \end{equation}
  As an auxiliary claim, we contend that for any $\tau\leq u/2$ there
  exists a $2\tau$-subset of $\cS\subseteq\Gamma$ that satisfies one
  of the following properties:
  \begin{itemize}
  \item[P1.] There exists a subset $\cS'\subseteq \cS$ and $S'\in
    \cS'$ such that
    \begin{equation}
      \label{eqn_P1}
      \abs{S'}-\delta+1\leq\abs{S'\cap \parenv{\bigcup_{S\in \cS'\setminus\mathset{S'}}S}}<\abs{S'}.
    \end{equation}
    \item[P2.] The following inequalities hold:
    \begin{align}
    &\abs{\cS}(r+\delta-1)-\abs{\bigcup_{S\in \cS}S}\geq \tau, \label{eqn_P2_1} \\
    &\abs{\bigcup_{S\in \cS}S}\geq \rank\parenv{\bigcup_{S\in \cS}S}+\abs{\cS}(\delta-1). \label{eqn_P2_2}
    \end{align}
  \end{itemize}
  We proceed to prove this auxiliary claim by induction on $\tau$.

  For the induction base, consider $\tau=1\leq u/2$. In that case,
  choose $\cS=\cS'=\mathset{\hS,\hS+\hatt}$. By~\eqref{eq:contrar}, if
  additionally, $|\hS|-\delta+1\leq |\hS\cap(\hS+\hatt)|$, then P1
  holds. Otherwise, by Lemma~\ref{lemma_for_rank_B_i}, P2 holds. Thus,
  the induction base is proved. Now arbitrarily choose $i_1\in
  \hS\cap(\hS+\hatt)$.

  For the induction hypothesis, assume the claim holds for $\tau$, and
  let $\cS_{\tau}$ be a set that satisfies the claim in that case,
  i.e., $\abs{\cS_{\tau}}=2\tau$. For the induction step, we prove it
  also holds for $\tau+1$, as long as $\tau+1\leq u/2$, namely, that
  there exists a repair set of repair sets, $\cS_{\tau+1}$, containing
  $2(\tau+1)$ repair sets, that satisfies P1 or P2. We shall make an
  educated guess as to what $\cS_{\tau+1}$ might be, which will work
  in most cases. When it does not, we shall offer a correction to our
  initial choice of $\cS_{\tau+1}$.

  Since
  $2\tau\leq u-2$ we have
  \[\rank\parenv{\bigcup_{S\in \cS_{\tau}}S}\leq 2\tau  r \leq (u-2)r < k.\]
  Hence, there exists an
  $i_{\tau+1}\in\Z_n$ with $\spn(\mathset{i_{\tau+1}})\not\subseteq
  \spn(\bigcup_{S\in \cS_{\tau}}S)$. As our initial guess, we now
  define the following:
  \begin{align*}
    S_{\tau+1,1}&=\hS+i_{\tau+1}-i_1, \\
    S_{\tau+1,2}&=\hS+\hatt+i_{\tau+1}-i_1,\\
    \cS_{\tau+1}&=\cS_{\tau}\cup\mathset{S_{\tau+1,1},S_{\tau+1,2}}.
  \end{align*}
  We observe that $S_{\tau+1,1}\neq S_{\tau+1,2}$ since they are
  cyclic rotations by the same amount of $\hS$ and $\hS+\hatt$,
  respectively, which by~\eqref{eq:contrar}, are two distinct sets.
  Additionally, $i_{\tau+1}\in S_{\tau+1,1}\cap S_{\tau+1,2}$, and
  since $\spn(\mathset{i_{\tau+1}})\not\subseteq \spn(\bigcup_{S\in
    \cS_{\tau}}S)$, it follows that $S_{\tau+1,1},S_{\tau+1,2}\not\in
  \cS_{\tau}$. Hence, $\abs{\cS_{\tau+1}}=2(\tau+1)$.

  If $\cS_{\tau}$ satisfies P1 then trivially so does $\cS_{\tau+1}$
  and the claim follows. Assume then that $S_{\tau}$ only satisfies
  P2. In particular, by~\eqref{eqn_P2_2},
  \begin{equation}
    \label{eq:onlyp2}
  \abs{\bigcup_{S\in \cS_{\tau}}S}\geq \rank\parenv{\bigcup_{S\in \cS_{\tau}}S}+\abs{\cS_{\tau}}(\delta-1).
  \end{equation}
  Again, if $\cS_{\tau+1}$ satisfies P1 then we are done. Otherwise,
  assume that $\cS_{\tau+1}$ does not satisfy P1, which means
  \begin{equation}\label{eqn_cover}
    \abs{S_{\tau+1,j}\cap \parenv{\bigcup_{S\in\cS_{\tau+1}\setminus\mathset{S_{\tau+1,j}}}S}}=\abs{S_{\tau+1,j}},
  \end{equation}
  or
  \begin{equation}\label{eqn_size_R}
    \abs{S_{\tau+1,j}\cap \parenv{\bigcup_{S\in \cS_{\tau+1}\setminus\mathset{S_{\tau+1,j}}}S}}<\abs{S_{\tau+1,j}}-\delta+1
  \end{equation}
  for $j=1,2$.

  If~\eqref{eqn_cover} holds for $S_{\tau+1,1}$, then the fact that
  \[0<\abs{S_{\tau+1,1}\cap S_{\tau+1,2}}=\abs{\hS\cap (\hS+\hatt)}< \abs{\hS},\]
  means that
  \begin{equation}
    \label{eq:atleastone}
    \abs{S_{\tau+1,1}\cap \parenv{\bigcup_{S\in\cS_{\tau}}S}}\geq 1.
  \end{equation}
  Recall that $\spn(\mathset{i_{\tau+1}})\not\subseteq
  \spn(\cup_{S\in\cS_{\tau}}S)$, but note that $i_{\tau+1}\in
  S_{\tau+1,1}$, which implies that
  \[\abs{S_{\tau+1,1}\cap \parenv{\bigcup_{S\in\cS_{\tau}}S}}<\abs{S_{\tau+1,1}}-\delta+1.\]
  Thus, we can find a $(\delta-1)$-subset $S^*_{\tau+1,1}\subseteq
  S_{\tau+1,1}$ such that $\rank(S_{\tau+1,1}\setminus
  S^*_{\tau+1,1})=\rank(S_{\tau+1,1})$ and
  \[S^*_{\tau+1,1}\cap \parenv{\bigcup_{S\in\cS_{\tau}}S}=\emptyset.\]
  We therefore have,
  \begin{equation}
    \label{eq:step1}
    \begin{split}
  \rank\parenv{S_{\tau+1,1}\cup\parenv{\bigcup_{S\in \cS_{\tau}}S}}&=\rank\parenv{(S_{\tau+1,1}\setminus S^*_{\tau+1,1})\cup\parenv{\bigcup_{S\in \cS_{\tau}}S}}\\
  &\leq\abs{S_{\tau+1,1}\setminus \parenv{S^*_{\tau+1,1}\cup \parenv{\bigcup_{S\in \cS_{\tau}}S}}}+\rank\parenv{\bigcup_{S\in \cS_{\tau}}S}\\
  &\leq \abs{S_{\tau+1,1}\setminus \parenv{\bigcup_{S\in \cS_{\tau}}S}}-\delta+1+\abs{\bigcup_{S\in \cS_{\tau}}S}-2\tau(\delta-1)\\
  &= \abs{S_{\tau+1,1}\cup\parenv{\bigcup_{S\in \cS_{\tau}}S}}-(2\tau+1)(\delta-1),\\
    \end{split}
  \end{equation}
  where the second inequality holds by~\eqref{eq:onlyp2}.
  Note that since $\cS_{\tau}$ satisfies P2, by~\eqref{eqn_P2_1},
  \begin{equation}
    \label{eq:step2}
  (2\tau+1)(r+\delta-1)\geq \abs{\bigcup_{S\in \cS_{\tau}}S}+\tau+r+\delta-1
  \geq \abs{\bigcup_{S\in \cS_{\tau}}S}+\tau+\abs{S_{\tau+1,1}}\geq \abs{S_{\tau+1,1}\cup \parenv{\bigcup_{S\in \cS_{\tau}}S}}+\tau+1,
  \end{equation}
  where the last inequality follows from~\eqref{eq:atleastone}. Recall that
  $\tau+1\leq u/2$, hence
  \[ \rank\parenv{S_{\tau+1,1}\cup\parenv{\bigcup_{S\in\cS_{\tau}} S}} \leq (2\tau+1)r\leq (u-1)r \leq k-1-r.\]
  It then follows that there exists a repair set
  $\tS_{\tau+1,2}\in\Gamma$ such that
  \[ \spn(\tS_{\tau+1,2}) \not\subseteq \spn\parenv{\bigcup_{S\in\cS_{\tau}\cup\mathset{S_{\tau+1,1}}}S}.\]
  We now correct our initial guess, and for this case only, set
  $\cS_{\tau+1}=\cS_{\tau}\cup\mathset{S_{\tau+1,1},\tS_{\tau+1,2}}$.
  We therefore have,
  \[\rank\parenv{\bigcup_{S\in\cS_{\tau+1}}S}>\rank\parenv{\bigcup_{S\in \cS_{\tau}\cup \mathset{S_{\tau+1,1}}}S}.\]
  By the last inequality, there exists a $(\delta-1)$-subset
  $\tS^*_{\tau+1,2}\subseteq \tS_{\tau+1,2}\setminus
  (\bigcup_{S\in\cS_{\tau}\cup\mathset{S_{\tau+1,1}}}S)$, and then
  \begin{equation}
    \label{eq:step3}
    \begin{split}
  \rank\parenv{\bigcup_{S\in \cS_{\tau+1}}S}&=\rank\parenv{(\tS_{\tau+1,2}\setminus \tS^*_{\tau+1,2})\cup\parenv{\bigcup_{S\in \cS_{\tau}\cup\mathset{S_{\tau+1,1}}}S}}\\
  &\leq\abs{\tS_{\tau+1,2}\setminus \parenv{\tS^*_{\tau+1,2}\cup \parenv{\bigcup_{S\in \cS_{\tau}\cup\mathset{S_{\tau+1,1}}}S}}}+\rank\parenv{\bigcup_{S\in \cS_{\tau}\cup\mathset{S_{\tau+1,1}}}S}\\
  &\leq \abs{\tS_{\tau+1,2}\setminus \parenv{\bigcup_{S\in \cS_{\tau}\cup\mathset{S_{\tau+1,1}}}S}}-\delta+1+\abs{S_{\tau+1,1}\cup\bigcup_{S\in \cS_{\tau}}S}-(2\tau+1)(\delta-1)\\
  &= \abs{\bigcup_{S\in \cS_{\tau+1}}S}-(2\tau+2)(\delta-1),\\
    \end{split}
  \end{equation}
  where the second inequality holds by~\eqref{eq:step1}.
  By~\eqref{eq:step2} we have,
  \begin{equation}
    \label{eq:step4}
    \abs{\cS_{\tau+1}}(r+\delta-1)-\abs{\bigcup_{S\in\cS_{\tau+1}}S}
    \geq (2\tau+1)(r+\delta-1)-\abs{S_{\tau+1,1}\cup\parenv{\bigcup_{S\in\cS_{\tau}}S}} \geq \tau+1.
  \end{equation}
  In total, the combination of~\eqref{eq:step3} and~\eqref{eq:step4} shows
  that the modified $\cS_{\tau+1}$ satisfies P2.

  We now return to the original
  $\cS_{\tau+1}=\cS_{\tau}\cup\mathset{S_{\tau+1,1},S_{\tau+1,2}}$. If
  $S_{\tau+1,2}$ satisfies~\eqref{eqn_cover}, then a similar argument
  shows we can build a modified $\cS_{\tau+1}$ for which P2 holds.

  As a final case, we consider the situation where both $S_{\tau+1,1}$
  and $S_{\tau+1,2}$ satisfy~\eqref{eqn_size_R}. In that case, there
  exist $(\delta-1)$-subsets $S^*_{\tau+1,j}\subseteq S_{\tau+1,j}$
  with $S^*_{\tau+1,j}\cap (\bigcup_{S\in
    \cS_{\tau+1}\setminus\mathset{S^*_{\tau+1,j}}}S)=\emptyset$ and
  $\rank(S_{\tau+1,j})=\rank(S_{\tau+1,j}\setminus S^*_{\tau+1,j})$,
  for $j=1,2$. Thus, we have
  \begin{equation}
    \label{eq:step7}
  \begin{split}
  \rank\parenv{\bigcup_{S\in \cS_{\tau+1}}S}&=\rank\parenv{(S_{\tau+1,1}\setminus S^*_{\tau+1,1})\cup (S_{\tau+1,2}\setminus S^*_{\tau+1,2})\cup\parenv{\bigcup_{S\in \cS_{\tau}}S}}\\
  &\leq \abs{((S_{\tau+1,1}\setminus S^*_{\tau+1,1})\cup (S_{\tau+1,2}\setminus S^*_{\tau+1,2}))\setminus \parenv{\bigcup_{S\in \cS_{\tau}}S}}+\rank\parenv{\bigcup_{S\in \cS_{\tau}}S}\\
  &\leq \abs{(S_{\tau+1,1}\cup S_{\tau+1,2})\setminus \parenv{\bigcup_{S\in \cS_{\tau}}S}}-2(\delta-1)+\rank\parenv{\bigcup_{S\in \cS_{\tau}}S}\\
  &\leq \abs{\bigcup_{S\in \cS_{\tau+1}}S}-(2\tau+2)(\delta-1),\\
  \end{split}
  \end{equation}
  where the last inequality holds by~\eqref{eqn_P2_2}. Additionally, by~\eqref{eqn_P2_1}, and since $i_{\tau+1}\in S_{\tau+1,1}\cap S_{\tau+1,2}$,
  \begin{equation}
    \label{eq:step8}
  2(\tau+1)(r+\delta-1)-\abs{\bigcup_{S\in \cS_{\tau+1}}S}\geq 2\tau(r+\delta-1)-
  \abs{\bigcup_{S\in \cS_{\tau}}S}+2(r+\delta-1)-\abs{S_{\tau+1,1}\cup S_{\tau+1,2}}\geq \tau+1.
  \end{equation}
  By combining~\eqref{eq:step7} and~\eqref{eq:step8} we learn that
  $\cS_{\tau+1}$ satisfies P2, and the auxiliary claim follows.

  We turn to prove the main claim. The proof is divided into two cases
  depending on properties P1 and P2:

  \textbf{Case 1:} P1 holds for some $\tau\leq \frac{u}{2}$. Let
  $\cS\subseteq\Gamma$ be a $2\tau$-subset, $\cS'\subseteq \cS$, and
  $S'\in \cS'$, such that~\eqref{eqn_P1} holds. By that equation, we
  can choose a subset $\cV\subseteq\cS'\setminus\mathset{S'}$ such
  that $\rank(\bigcup_{S\in \cV}S)=\rank(\bigcup_{S\in \cS'}S)$. Of
  all such subsets, let us choose $\cV$ to be minimal, namely,
  $\rank(\bigcup_{S\in \cV}S)>\rank(\bigcup_{S\in \cV\setminus
    \mathset{A}}S)$ for any $A\in \cV$.  Thus, by
  Lemma~\ref{lemma_for_rank_B_i} we have
  \begin{equation}\label{eqn_rank_V}
  \rank\parenv{\bigcup_{S\in \cV}S}\leq \abs{\bigcup_{S\in \cV}S}-\abs{\cV}(\delta-1).
  \end{equation}
  Assume $\cV$ contains $\nu$ repairs sets,
  $\cV=\mathset{S_1,S_2,\dots,S_\nu}$. Since each repair set in
  $\Gamma$ has rank at most $r$, and the union of $u$ repair sets has
  rank at most $ur\leq k-1$, we can extend $\cV$ to a $u$-set
  $\cV'\subseteq\Gamma$ such that
  $\cV'=\cV\cup\mathset{S_{\nu+1},S_{\nu+2},\dots, S_{u}}$, such that
  each added repair set increases the overall rank, i.e.,
  \begin{equation}\label{eqn_rank_V_1_first}
  \rank\parenv{\cV\cup\mathset{S_{\nu+1},S_{\nu+2},\dots, S_{\nu+i}}}<\rank\parenv{\cV\cup\mathset{S_{\nu+1},S_{\nu+2},\dots, S_{\nu+i+1}}}
  \end{equation}
  for all $1\leq i\leq u-\nu-1$.

  Let $S^*_{\nu+1}$ be a $(\delta-1)$-subset of $S_{\nu+1}\setminus
  (\bigcup_{S\in \cV}S)$ and $\overline{S'}\eqdef S'\setminus
  (\bigcup_{S\in \cV}S)$. In a similar fashion to the analysis above,
  we have
  \begin{align*}
  \rank\parenv{\bigcup_{S\in \cV\cup\{S_{\nu+1}\}}S}&=\rank\parenv{(S_{\nu+1}\setminus S^*_{\nu+1})\cup \parenv{\bigcup_{S\in \cV}S}}\\
  &\leq
  \begin{cases}
  |(S_{\nu+1}\setminus S^*_{\nu+1})\setminus (\bigcup_{S\in \cV}S)|-1+\rank(\bigcup_{S\in \cV}S),& \text{ if } \overline{S'}\cap S_{\nu+1}\ne \emptyset\\
  |(S_{\nu+1}\setminus S^*_{\nu+1})\setminus (\bigcup_{S\in \cV}S)|+\rank(\bigcup_{S\in \cV}S),& \text{ otherwise}\\
  \end{cases}\\
  &\leq
  \begin{cases}
  |S_{\nu+1}\setminus (\bigcup_{S\in \cV}S)|-\delta+\rank(\bigcup_{S\in \cV}S),& \text{ if } \overline{S'}\cap S_{\nu+1}\ne \emptyset\\
  |S_{\nu+1}\setminus (\bigcup_{S\in \cV}S)|-\delta+1+\rank(\bigcup_{S\in \cV}S),& \text{ otherwise}\\
  \end{cases}\\
  & \leq
  \begin{cases}
  |S_{\nu+1}\setminus (\bigcup_{S\in \cV}S)|-\delta+|\bigcup_{S\in \cV}S|-|\cV|(\delta-1),& \text{ if } \overline{S'}\cap S_{\nu+1}\ne \emptyset\\
  |S_{\nu+1}\setminus (\bigcup_{S\in \cV}S)|-\delta+1+|\bigcup_{S\in \cV}S|-|\cV|(\delta-1),& \text{ otherwise}\\
  \end{cases}\\
  &\leq
  \begin{cases}
  |\bigcup_{S\in \cV\cup\{S_{\nu+1}\}}S|-(|\cV|+1)(\delta-1)-1,& \text{ if } \overline{S'}\cap S_{\nu+1}\ne \emptyset\\
  |\bigcup_{S\in \cV\cup\{S_{\nu+1}\}}S|-(|\cV|+1)(\delta-1),& \text{ otherwise}\\
  \end{cases}
  \end{align*}
  where to prove the first inequality we use the fact that
  $S'\subseteq \spn(\bigcup_{S\in \cV}S)$. Repeating the processing,
  at each iteration adding $S_{\nu+2},\dots,S_u$, we can conclude that
  \begin{equation}\label{eqn_rank_V_1}
  \rank\parenv{\bigcup_{S\in \cV'}S}\leq
  \begin{cases}
  |\bigcup_{S\in \cV'}S|-u(\delta-1)-1,& \text{ if } \overline{S'}\cap (\bigcup_{1\leq i\leq u-\nu-1}S_{\nu+i})\ne\emptyset\\
  |\bigcup_{S\in \cV'}S|-u(\delta-1),&\text{ otherwise,}\\
  \end{cases}
  \end{equation}
  by \eqref{eqn_rank_V} and \eqref{eqn_rank_V_1_first}.

  Recall that the rank of the union of $u$ repair sets, and in
  particular, $\cV'$, satisfies $\rank(\bigcup_{S\in \cV'}S)\leq
  ur\leq k-1$. Thus, we have a set of coordinates $B\subseteq\Z_n$,
  with $\bigcup_{S\in \cV'}S\subseteq B$ and $\rank(B)=k-1$. Consider
  the set $\widetilde{B}\eqdef B\cup \overline{S'}$. By
  \eqref{eqn_rank_V_1},
  \begin{align*}
    \abs{\widetilde{B}}-\rank\parenv{\widetilde{B}}=\abs{\widetilde{B}}-\rank(B)
    &\geq
    \begin{cases}
      |B|-\rank(B), &\text{ if } \overline{S'}\cap (\bigcup_{1\leq i\leq u-\nu-1}S_{\nu+i})\ne\emptyset\\
      |B|-\rank(B)+1, &\text{ otherwise}
    \end{cases}\\
    &\geq
    \begin{cases}
      |\bigcup_{S\in \cV'}S|-\rank(\bigcup_{S\in \cV'}S), &\text{ if } \overline{S'}\cap (\bigcup_{1\leq i\leq u-\nu-1}S_{\nu+i})\ne\emptyset\\
      |\bigcup_{S\in \cV'}S|-\rank(\bigcup_{S\in \cV'}S)+1, &\text{ otherwise}
    \end{cases}\\
    &\geq u(\delta-1)+1,
  \end{align*}
  i.e.,
  \begin{equation}
    \label{eq:btilde}
    \abs{\widetilde{B}}\geq k+u(\delta-1).
  \end{equation}
  Recall now that for an $[n,k,d]_q$ code $\cC$,
  \[ d = n- \max\mathset{\abs{I} ~:~ I\subseteq\Z_n, \rank(\cC_I)=k-1 }.\]
  Thus, by~\eqref{eq:btilde}, for our code
  \[ d \leq n-k-u(\delta-1).\]
  However, since our code is an optimal LRC,
  \[ d = n-k+1-u(\delta-1),\]
  and thus, a have reached a contradiction.

  \textbf{Case 2:} P2 holds for all $2\tau$-subsets
  $\cS\subseteq\Gamma$, where $\tau\leq u/2$. Assume first that $u$ is
  odd. Denote $\tau=\frac{u-1}{2}$, and arbitrarily pick
  $\cS\subseteq\Gamma$, with
  $\abs{\cS}=2\tau=u-1$. By~\eqref{eqn_P2_1} and~\eqref{eqn_P2_2},
  \[
    k-1-\rank\parenv{\bigcup_{S\in \cS}S}=ur+v-1-\rank\parenv{\bigcup_{S\in \cS}S} \geq r+v-1+\frac{u-1}{2}\geq 2r,
  \]
  where the last inequality holds by the condition $u\geq 2(r-v+1)$,
  and the fact that $u$ is odd. Thus, we can extend $\cS$ to
  $\cV'=\cS\cup \mathset{S_{u}, S_{u+1}}\subseteq \Gamma$ with
  $\abs{\cV'}=u+1$, such that
  \[\rank\parenv{\bigcup_{S\in \cS}S}<\rank\parenv{S_{u}\cup\parenv{\bigcup_{S\in\cS}S}}<\rank\parenv{\bigcup_{S\in \cV'}S}\leq k-1,\]
  and
  \[\abs{\bigcup_{S\in \cV'}S}\geq \rank\parenv{\bigcup_{S\in \cV'}S}+(u+1)(\delta-1).\]
  The fact that $\rank(\bigcup_{S\in \cV'}S)\leq k-1$ means that we can
  find a set $B\subseteq\Z_n$ with $\bigcup_{S\in \cV'}S\subseteq B$
  and $\rank(B)=k-1$. Then,
  \[|B|-k+1=|B|-\rank(B)\geq \abs{\bigcup_{S\in \cV'}S}-\rank\parenv{\bigcup_{S\in \cV'}S}\geq (u+1)(\delta-1).\]
  As in Case 1, we obtain
  \[ d\leq n-k+1-(u+1)(\delta-1),\]
  which contradicts the minimum distance of an optimal LRC being
  \[ d=n-k+1-u(\delta-1).\]

  Assume now that $u$ is even. Denote $\tau=\frac{u}{2}$, and
  arbitrarily pick $\cS\subseteq\Gamma$, with
  $\abs{\cS}=2\tau=u$. By~\eqref{eqn_P2_1} and~\eqref{eqn_P2_2},
  \[
    k-1-\rank\parenv{\bigcup_{S\in \cS}S}=ur+v-1-\rank\parenv{\bigcup_{S\in \cS}S} \geq v-1+\frac{u}{2}\geq r,
  \]
  where the last inequality holds by the condition $u\geq
  2(r-v+1)$. Thus, we can extend $\cS$ to $\cV'=\cS\cup
  \mathset{S_{u+1}}\subseteq \Gamma$ with $\abs{\cV'}=u+1$, such that
  \[\rank\parenv{\bigcup_{S\in \cS}S}<\rank\parenv{\bigcup_{S\in \cV'}S}\leq k-1,\]
  and
  \[\abs{\bigcup_{S\in \cV'}S}\geq \rank\parenv{\bigcup_{S\in \cV'}S}+(u+1)(\delta-1).\]
  We now continue exactly as in the case of odd $u$ to obtain a contradiction.

  In all of the above cases, we have reached a contradiction. Hence,
  our assumption that there exist $\hS\in\Gamma$ and $\hatt\in\Z$ such
  that $0<|\hS\cap (\hS+\hatt)|<|\hS|$ is incorrect, and the main
  claim of the theorem follows.
\end{IEEEproof}

\bibliographystyle{IEEEtranS}
\bibliography{HanBib}

\end{document}